\documentclass[english, 10pt, letterpaper]{article} 
\usepackage{tcolorbox} 
\usepackage{xcolor} 
\definecolor{myred}{RGB}{153, 0, 0} 
\definecolor{mygreen}{RGB}{0, 70, 0} 
\definecolor{myblue}{RGB}{0, 102, 204} 
\usepackage{cite} 
\usepackage[T1]{fontenc} 
\usepackage[left = 0.59in, right = 0.59in, top = 0.89in, bottom = 0.72in]{geometry} 
\usepackage{wrapfig} 
\usepackage{graphicx} 
\usepackage{authblk} 
\usepackage[hidelinks]{hyperref} 
\usepackage{orcidlink} 
\usepackage{balance} 
\usepackage[misc]{ifsym} 
\usepackage{amssymb} 
\usepackage{amsmath} 
\usepackage{gensymb} 
\usepackage{babel} 
\usepackage{indentfirst} 
\usepackage{setspace} 
\usepackage{lettrine} 
\usepackage{fancyhdr} 
\usepackage{stackengine} 
\usepackage{caption} 
\providecommand{\indexterms}[1]{\small{\textbf{\textit{Index Terms---}}}#1} 
\providecommand{\abst}[1]{\textbf{\textit{Abstract---}}#1} 
\usepackage[scaled = 1]{helvet} 
\renewcommand\cite[1]{{\color{mygreen}[\citenum{#1}]}} 
\title{\Huge\textbf{Oscillation with Negative Impedance}}
\author{\textbf{Taeju Lee}{\,}\orcidlink{0000-0001-6304-3073}}
\affil{\small{Department of Electrical Engineering, Columbia University, New York, NY 10027, USA}\\
\small{{\Letter}{\,}e-mail: taeju.leo.lee@gmail.com}}
\date{} 

\begin{document}
\onecolumn 
\maketitle
\pagestyle{fancy}
\fancyhf{}
\lhead{\footnotesize{LEE: OSCILLATION WITH NEGATIVE IMPEDANCE}}
\cfoot{\thepage} 

\begin{center}
\begin{tcolorbox}[colframe = black!0!white, colback = black!0!white, coltext = black, height = 7cm, width = 14cm, after = \vspace{0pt}]
\noindent\small{\abst{\textbf{Oscillation and frequency modulation have been leveraged for applications in detection (or sensing), data processing, and telemetry. This work provides a theoretical analysis of oscillation phenomena with negative impedance implemented using a cross-coupled transconductance pair. The negative impedance can consist not only of a negative resistance but also of a negative inductance and a negative capacitance, where the negative resistance supplies energy to the inductance and capacitance, causing oscillation at a resonant frequency. Also, the resonant frequency can be modulated by combining with passive components such as a resistor, an inductor, and a capacitor. In this work, a comprehensive circuit analysis employing small-signal models and transfer functions is performed to understand the oscillation phenomena and resonant frequencies arising from a combination of active and passive RLC circuits, in which the active RLC circuit is modeled as a negative impedance and implemented using a cross-coupled transconductance pair.}}}

\singlespacing\noindent\small{\indexterms{\textbf{Cross-coupled pair, detection, data processing, frequency modulation, LC oscillator, negative impedance, negative resistance, negative inductance, negative capacitance, oscillation, positive feedback, telemetry, resonance, sensing.}}}
\end{tcolorbox}
\vfill{ 
\begin{center}
\begin{tcolorbox}[colframe = black!0!white, colback = black!0!white, coltext = black, height = 1cm, width = 18cm]
\begin{wrapfigure}{r}{0.1\textwidth}
\vspace{-1.5cm}\hspace{-8.15cm}\includegraphics[scale = 0.2]{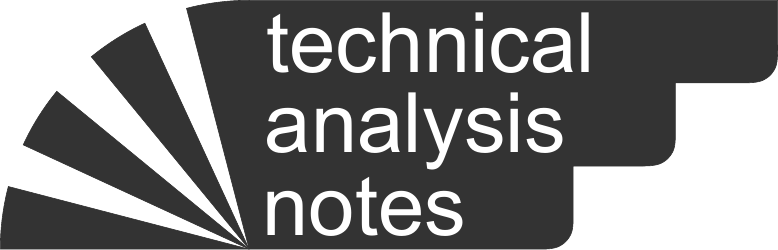}\label{logo}
\end{wrapfigure}
\noindent\rule{\textwidth}{1pt} 
\raggedleft\footnotesize{\textit{Technical Analysis Notes}{\,}\copyright{\,}2026 Taeju Lee. All Rights Reserved.} 
\end{tcolorbox}
\end{center}
} 
\end{center}

\rfoot{\footnotesize{\textit{Technical Analysis Notes}}} 
\newpage
\tableofcontents

\newpage
\twocolumn
\section{\color{myblue}\Large{I}\large{NTRODUCTION}}
\begin{figure}[h!]
\centering\includegraphics[scale = 0.55]{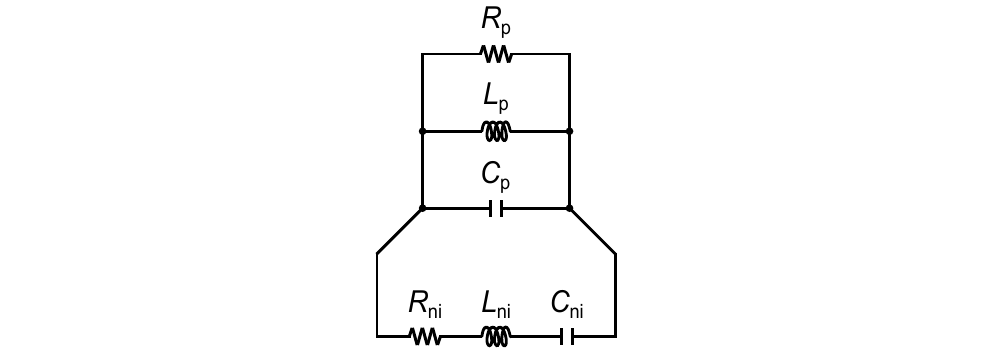}
\captionof{figure}{{\color{myred}\textbf{Oscillation model.}} A negative impedance combined with passive components in parallel.}\label{fig. 1}
\end{figure}

\lettrine[findent = 0pt, nindent = 0pt]{\textbf{O}}{\textbf{scillation}} and frequency modulation (FM) have been leveraged in various applications, including telemetry \cite{Chang1970OFDM}\text{\color{mygreen}--}\cite{Nissel2017MCM}, as well as detection \cite{Helmy2012detection}\text{\color{mygreen}--}\cite{Qiao2018detection} and data processing \cite{Leroux2021computing}\text{\color{mygreen}--}\cite{Ji2024computing}. This paper provides a comprehensive analysis of oscillation phenomena arising from an oscillation model (Fig. \ref{fig. 1}). The oscillation model consists of a series RLC circuit{\textemdash}$R_{ni}$, $L_{ni}$, and $C_{ni}${\textemdash}and a parallel circuit{\textemdash}$R_{p}$, $L_{p}$, and $C_{p}$, where the series RLC circuit is implemented using a cross-coupled transconductance pair. $R_{ni}$, $L_{ni}$, and $C_{ni}$ are called the "active resistance", the "active inductance", and the "active capacitance", respectively, where these are obtained as negative magnitudes and used to generate oscillation. The operating principles of these active components will be derived in later sections. Before delving into oscillation phenomena, basic principles of an RLC circuit are discussed in the following section.

\subsection{\color{myblue}Basic RLC Circuit}
\begin{figure}[h!]
\centering\includegraphics[scale = 0.55]{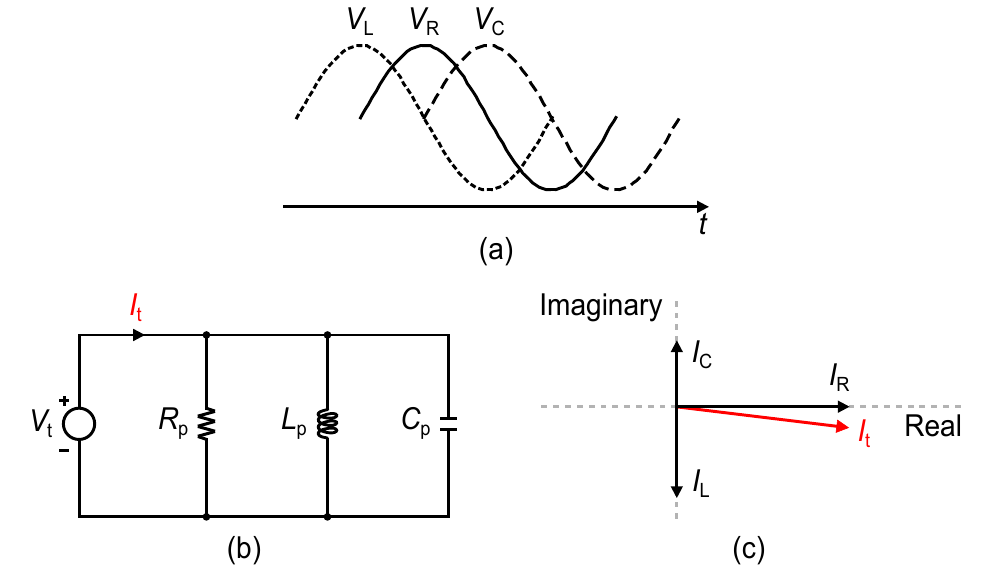}
\caption{{\color{myred}\textbf{RLC basic.}} (a) Transient response in a series RLC circuit. (b) Parallel RLC circuit. (c) Phasor diagram of a parallel RLC circuit.}\label{fig. 2}
\end{figure}

In a series RLC circuit, when a time-varying current $i(t)$ flows through a resistor $R_{s}$, a voltage drop $V_{R}$ across $R_{s}$ is $iR_{s}$ and has the same phase as the current, whereas a voltage drop $V_{L}$ across an inductor $L_{s}$ is $L_{s}{\cdot}(di/dt)$ with a $-90^{\circ}$ phase shift compared to $V_{R}$, and a voltage drop $V_{C}$ across a capacitor $C_{s}$ is expressed as $(1/C_{s}){\cdot}\int{i(t){\,}dt}$ with a $+90^{\circ}$ phase shift compared to $V_{R}$ (Fig. \ref{fig. 2}(a)). Accordingly, a phase difference between $V_{L}$ and $V_{C}$ is $180^{\circ}$.

To analyze oscillation phenomena, a series RLC circuit is employed, along with a parallel RLC circuit consisting of a resistor $R_{p}$, an inductor $L_{p}$, and a capacitor $C_{p}$ (Fig. \ref{fig. 2}(b)). \footnote{These parallel components{\textemdash}$R_{p}$, $L_{p}$, and $C_{p}${\textemdash}are the same as the passive components shown in Fig. \ref{fig. 1}.} To understand their impedances, a test voltage $V_{t}$ is applied to the parallel RLC circuit, and the resulting current $I_{t}$ is observed (Fig. \ref{fig. 2}(b)). $I_{t}$ can be expressed by summing the admittances of the parallel components in the phasor diagram (Fig. \ref{fig. 2}(c)). \footnote{Since the magnitudes of $-j({\omega}L_{p})^{-1}$ and $j{\omega}C_{p}$ are not equal, their phasors are not canceled out, and $I_{t}$ has a phase angle in Fig. \ref{fig. 2}(c).} The admittances of $L_{p}$ and $C_{p}$ are expressed as $-j({\omega}L_{p})^{-1}$ and $j{\omega}C_{p}$, respectively, resulting in a $180^{\circ}$ phase difference as in Fig. \ref{fig. 2}(c).

\subsection{\color{myblue}Parallel Resonance}
\begin{figure}[h!]
\centering\includegraphics[scale = 0.55]{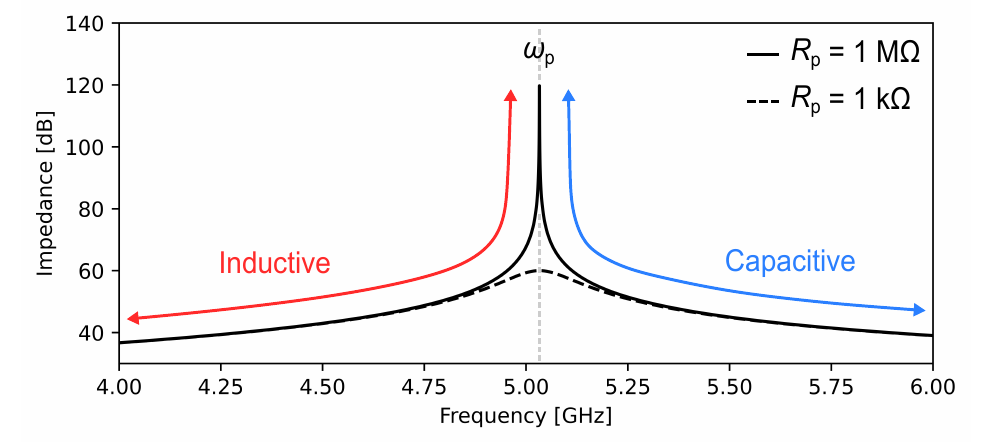}
\caption{{\color{myred}\textbf{Frequency response.}} Impedance magnitude of a parallel RLC circuit with $L_{p}=1{\,}\mathrm{nH}$ and $C_{p}=1{\,}\mathrm{pF}$.}\label{fig. 3}
\end{figure}

The impedance $Z_{p}$ of the parallel RLC circuit is obtained by $V_{t}/I_{t}$ (Fig. \ref{fig. 2}(b)), which is $R_{p}||j{\omega}L_{p}||(j{\omega}C_{p})^{-1}$. The $Z_{p}$ magnitude is dominated by a low-impedance path among the components. Therefore, assuming that $R_{p}{\,}{\gg}{\,}L_{p}{\,}{\gg}{\,}C_{p}$, $Z_{p}$ is dominated by $j{\omega}L_{p}$ in a relatively low-frequency range, whereas $Z_{p}$ is dominated by $(j{\omega}C_{p})^{-1}$ in a relatively high-frequency range (Fig. \ref{fig. 3}).

If the impedance magnitudes of $L_{p}$ and $C_{p}$ are equal by ${\omega}L_{p}=({\omega}C_{p})^{-1}$, the currents flowing through the two components are canceled out at a parallel resonant frequency $f_{p}$ because the phase difference of the two currents is $180^{\circ}$ (Fig. \ref{fig. 2}(c)). This means that the effective current flowing through the $L_{p}$ and $C_{p}$ becomes zero at $f_{p}$. Consequently, the impedance seen looking into the parallel-connected $L_{p}$ and $C_{p}$ becomes infinite because no current flows through the $L_{p}$ and $C_{p}$ $(j{\omega}L_{p}||(j{\omega}C_{p})^{-1}=\infty)$. According to infinity of $j{\omega}L_{p}||(j{\omega}C_{p})^{-1}$ at $f_{p}$, the impedance $Z_{p}$, given by $V_{t}/I_{t}$, is approximated as $R_{p}$ at $f_{p}$ (Fig. \ref{fig. 3}), meaning that a higher value of $R_{p}$ leads to an improved quality factor $Q$. Therefore, applying $V_{t}$ to $Z_{p}$ and observing $I_{t}$, $Z_{p}$ reaches its maximum value $R_{p}$ at $f_{p}$, where $I_{t}$ is observed as a minimum current $i_{p,min}$ flowing through $R_{p}$.

The impedance profile of $R_{p}||j{\omega}L_{p}||(j{\omega}C_{p})^{-1}$ is shown in Fig. \ref{fig. 3}, which exhibits inductive and capacitive characteristics depending on the frequency region, and demonstrates that a higher value of $R_{p}$ results in an improved $Q$. By solving ${\omega}_{p}=2{\pi}f_{p}$ and ${\omega}_{p}L_{p}=({\omega}_{p}C_{p})^{-1}$, where the current phasors of $L_{p}$ and $C_{p}$ are canceled out, the parallel resonant frequency $f_{p}$ is expressed as $1/(2{\pi}\sqrt{L_{p}C_{p}})$, which will be used to describe an oscillation principle in the following sections.

\subsection{\color{myblue}Series Resonance}
Compared to the parallel RLC circuit analyzed using the current phasors as in Fig. \ref{fig. 2}(c), a series RLC circuit can be analyzed using voltage phasors. Assuming that a series RLC circuit consisting of a resistor $R_{s}$, an inductor $L_{s}$, and a capacitor $C_{s}$, the impedance $Z_{s}$ of the series RLC circuit is $R_{s}+j{\omega}L_{s}+(j{\omega}C_{s})^{-1}$, where each component shows a different phase shift in the voltage domain while exhibiting the same phase shift in the current domain (Fig. \ref{fig. 2}(a)). Therefore, a voltage phasor (denoted by "impedance") is used in a series RLC circuit, and a current phasor (denoted by "admittance") is used in a parallel RLC circuit. \footnote{In the series RLC circuit, the voltage phasors of $L_{s}$ and $C_{s}$ are $j{\omega}L_{s}$ and $-j({\omega}C_{s})^{-1}$, respectively. In the parallel RLC circuit, the current phasors of $L_{p}$ and $C_{p}$ are $-j({\omega}L_{p})^{-1}$ and $j{\omega}C_{p}$, respectively.}

Performing a frequency sweep, since the magnitude of $Z_{s}$ is dominated by a high-impedance component, $Z_{s}$ is dominated by $(j{\omega}C_{s})^{-1}$ in a relatively low-frequency range, and $Z_{s}$ is dominated by $j{\omega}L_{s}$ in a relatively high-frequency range. \footnote{An impedance of a parallel circuit is dominated by a low-impedance component, whereas an impedance of a series circuit is dominated by a high-impedance component.} If the impedance magnitudes of $L_{s}$ and $C_{s}$ are equal by ${\omega}L_{s}=({\omega}C_{s})^{-1}$, where a series resonant frequency $f_{s}$ is obtained as $1/(2{\pi}\sqrt{L_{s}C_{s}})$, the voltage drops across $L_{s}$ and $C_{s}$ are canceled out because the voltage phasors of $L_{s}$ and $C_{s}$ show a $180^{\circ}$ phase difference in the series RLC circuit (Fig. \ref{fig. 2}(a)). This means that the effective impedance of the $L_{s}$ and $C_{s}$ becomes a short circuit (or zero impedance) because the voltage drop across the $L_{s}$ and $C_{s}$ becomes zero. Consequently, applying a test voltage $V_{t}$ to $Z_{s}$ and observing the resulting current $I_{t}$, $Z_{s}$ reaches its minimum value $R_{s}$ at $f_{s}$, where $I_{t}$ is observed as a maximum current $i_{s,max}$ flowing through $R_{s}$.

\subsection{\color{myblue}Summary of Resonant Circuits}
In summary, when sweeping from low to high frequencies, the parallel RLC circuit exhibits the impedance characteristics in the order of inductance, resonance with a maximum value of $R_{p}$, and capacitance (Fig. \ref{fig. 3}). Similarly, when sweeping from low to high frequencies in the series RLC circuit, the impedance characteristics appear in the order of capacitance, resonance with a minimum value of $R_{s}$, and inductance. As the parallel and series RLC circuits reach their resonant frequencies, $f_{p}$ (parallel resonance) and $f_{s}$ (series resonance), the impedance magnitude of the parallel RLC circuit reaches the maximum value of $R_{p}$ and the impedance magnitude of the series RLC circuit reaches the minimum value of $R_{s}$. As a result, the minimum current $i_{p,min}$ is drawn through $Z_{p}$ (${\approx}{\,}R_{p}$) at $f_{p}$, and the maximum current $i_{s,max}$ flows through $Z_{s}$ (${\approx}{\,}R_{s}$) at $f_{s}$. Note that all the components used in the above parallel and series RLC circuits are passive components with positive values. 
\section{\color{myblue}\Large{O}\large{SCILLATION} \Large{P}\large{RINCIPLE}}
To understand the oscillation principle, let us calculate the impedance $Z_{p}$ of the parallel RLC circuit consisting of $R_{p}$, $L_{p}$, and $C_{p}$ as depicted in Fig. \ref{fig. 2}(b). $Z_{p}$ is given by

\begin{align}
R_{p}||j{\omega}L_{p}||\frac{1}{j{\omega}C_{p}}&=R_{p}||j\left(\frac{{\omega}L_{p}}{1-{\omega}^2L_{p}C_{p}}\right)\\
&=\frac{j{\omega}R_{p}L_{p}}{R_{p}(1-{\omega}^2L_{p}C_{p})+j{\omega}L_{p}}
\end{align}

The above equation is decomposed into the resistance $R_{zp}$ and the reactance $X_{zp}$ as follows:

\begin{align}
Z_{p}=\underbrace{\frac{{\omega}^2R_{p}L_{p}^2}{R_{p}^2K_{p}^2+{\omega}^2L_{p}^2}}_{R_{zp}}+j\underbrace{\left(\frac{{\omega}R_{p}^2L_{p}K_{p}}{R_{p}^2K_{p}^2+{\omega}^2L_{p}^2}\right)}_{X_{zp}}
\end{align}
 
{\noindent}where $K_{p}=1-{\omega}^2L_{p}C_{p}$ and that is used to find the parallel resonant frequency $f_{p}$ (or ${\omega}_{p}=2{\pi}f_{p}$). In $K_{p}$ and $R_{zp}$, as ${\omega}$ increases from low to high values but does not exceed ${\omega}_{p}$, which is $1/\sqrt{L_{p}C_{p}}$, then $R_{zp}$ increases as $K_{p}$ decreases, showing an inductive behavior of $R_{zp}$ for ${\omega}<{\omega}_{p}$. As ${\omega}$ reaches ${\omega}_{p}$, $K_{p}$ becomes zero, $X_{zp}$ therefore vanishes and $R_{zp}$ achieves its maximum value $R_{p}$, causing the parallel RLC circuit to resonate. As ${\omega}$ increases beyond ${\omega}_{p}$, $K_{p}$ increases and $R_{zp}$ is therefore reduced, exhibiting a capacitive behavior of $R_{zp}$ for ${\omega}>{\omega}_{p}$. These inductive and capacitive behaviors are observed as depicted in Fig. \ref{fig. 3}.

\begin{figure}[h!]
\centering\includegraphics[scale = 0.55]{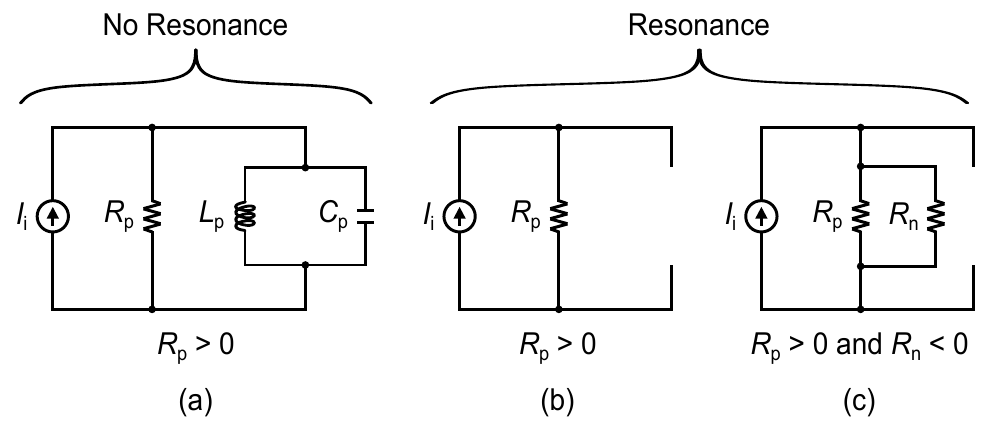}
\captionof{figure}{{\color{myred}\textbf{A parallel RLC circuit with a current impulse $I_{i}$.}} (a) No resonance as $R_{p}>0$. (b) Resonance as $R_{p}>0$. (c) Resonance as $R_{p}>0$ and $R_{n}<0$.}\label{fig. 4}
\end{figure}

Assuming that $R_{p}>0$ and a current impulse $I_{i}$ is connected with the parallel RLC circuit as shown in Fig. \ref{fig. 4}(a), $I_{i}$ is consumed by the components according to the inductive and capacitive impedance characteristics for ${\omega}<{\omega}_{p}$ and ${\omega}>{\omega}_{p}$, respectively. This is because the resistance $R_{zp}$ is positive as $R_{p}>0$ for ${\omega}<{\omega}_{p}$ and ${\omega}>{\omega}_{p}$. Note that a positive resistance dissipates energy, while a negative resistance is involved in energy generation \cite{Chua1987resistor}. When ${\omega}={\omega}_{p}$, the impedance of the $L_{p}$ and $C_{p}$ becomes infinite, resulting in an open circuit, leaving only $I_{i}$ and $R_{p}$ (Fig. \ref{fig. 4}(b)). Since $R_{p}$ is positive for ${\omega}={\omega}_{p}$, the power $I_{i}^2R_{p}$ is dissipated through $R_{p}$, occuring no oscillation at ${\omega}_{p}$ (or $f_{p}=1/(2{\pi}\sqrt{L_{p}C_{p}})$).

However, if the value of $R_{p}$ changes from positive to negative, oscillation at ${\omega}_{p}$ is revived because a negative resistance can supply energy compared to a positive resistance that absorbs (or dissipates) energy. Changing $R_{p}$ from positive to negative can be realized by connecting $R_{p}$ in parallel with a negative resistance $R_{n}$ (Fig. \ref{fig. 4}(c)). Assuming that the parallel resistance of $R_{p}$ and $R_{n}$ is given by $R_{pn}={R_{p}R_{n}}/(R_{p}+R_{n})$, the power $I_{i}^2R_{pn}$ at ${\omega}_{p}$ is not dissipated but continuously sustained as long as $R_{pn}<0$. For $R_{pn}$ to remain negative, the absolute value of $R_{n}$ must be less than that of $R_{p}$, meaning that $|R_{n}|<|R_{p}|$.

From another perspective, when $R_{n}$ is connected with $R_{p}$ in parallel while remaining $|R_{n}|<|R_{p}|$, the resistance of $Z_{p}$, which is $R_{zp}$, is also modified to $R_{zp}^{\prime}$ as follows:

\begin{align}
R_{zp}^{\prime}=\frac{{\omega}^2\left(\frac{R_{p}R_{n}}{R_{p}+R_{n}}\right)L_{p}^2}{\left(\frac{R_{p}R_{n}}{R_{p}+R_{n}}\right)^2K_{p}^2+{\omega}^2L_{p}^2}
\end{align}

{\noindent}where $R_{zp}^{\prime}$ remains negative, meaning that the power $I_{i}^2R_{zp}^{\prime}$ therefore remains negative over the entire frequency range. But the absolute value of $I_{i}^2R_{zp}^{\prime}$ is maximized at ${\omega}_{p}$, therefore, oscillation occurs at ${\omega}_{p}$, and oscillation does not occur for ${\omega}<{\omega}_{p}$ and ${\omega}>{\omega}_{p}$. In other words, the oscillation occurs at the point where $I_{i}^2R_{zp}^{\prime}$ is maximized in the negative direction over the entire frequency range, which is ${\omega}_{p}$ or $1/(2{\pi}\sqrt{L_{p}C_{p}})$. Note that assuming that $I_{i}$ is an ideal current source with an infinite output resistance in Fig. \ref{fig. 4}, the parallel combination of $I_{i}$ and $R_{p}$ can be modeled as a practical current source that consists of the magnitude $|I_{i}|$ and the output resistance $R_{p}$ ($I_{i}>0$ and $R_{p}>0$), so the load consisting of $L_{p}$ and $C_{p}$ is connected with the practical current source, which has the output resistance $R_{p}$.

Also, the above oscillation principle can be understood in another way. Assuming that ${\omega}L_{p}=({\omega}C_{p})^{-1}$ (Fig. \ref{fig. 2}(b)), the currents flowing through $L_{p}$ and $C_{p}$ are reciprocally canceled out in the phasor diagram, resulting in a net current of zero at the resonant frequency of $1/(2{\pi}\sqrt{L_{p}C_{p}})$. Therefore, the impedance seen looking into the $L_{p}$ and $C_{p}$ becomes infinite, and $Z_{p}$ is approximated as $R_{p}$. Under this condition, if $R_{p}$ is positive, energy at the resonant frequency is absorbed through $R_{p}$ of the parallel RLC circuit. Eventually, the circuit makes no oscillation at the resonant frequency.

However, if $R_{p}$ changes from positive to negative, the oscillation mechanism can be understood as follows: (1) Instead of absorbing energy from $L_{p}$ and $C_{p}$ to $R_{p}$, $R_{p}$ supplies energy to the remaining components, $L_{p}$ and $C_{p}$. (2) Over the entire frequency range, noise at the frequency $1/(2{\pi}\sqrt{L_{p}C_{p}})$ is triggered as an energy source. (3) This energy source is not dissipated but is continuously sustained by $R_{p}$ and supplied to $L_{p}$ and $C_{p}$, resulting in oscillation at the frequency $1/(2{\pi}\sqrt{L_{p}C_{p}})$.

Through the above mechanisms, oscillation can be generated at the resonant frequency by using the negative resistance. Assuming that a negative resistance $R_{n}$ is connected in parallel with $R_{p}$ (Fig. \ref{fig. 4}(c)), the parallel resistance expressed as ${R_{p}R_{n}}/(R_{p}+R_{n})$ always remains negative as long as $|R_{n}|<|R_{p}|$. The implementation of $R_{n}$ will be discussed in the following sections. 
\subsection{\color{myblue}Negative Impedance{\textemdash}Type $\mathrm{I}$}
\begin{figure}[h!]
\centering\includegraphics[scale = 0.55]{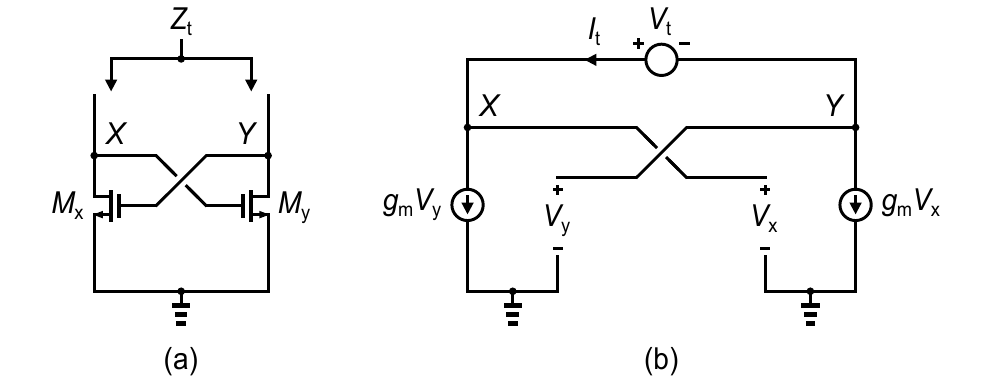}
\caption{{\color{myred}\textbf{Negative impedance{\textemdash}Type $\mathrm{I}$.}} (a) A cross-coupled transconductance pair. (b) Its small-signal model.}\label{fig. 5}
\end{figure}

The above negative resistance $R_{n}$ can be implemented by cross-coupling two transconductances. \footnote{Detailed information about a cross-coupled pair and its application can be found in \cite{Razavi2014cross_v1}\text{\color{mygreen}--}\cite{Razavi2015cross_v3}.} Fig. \ref{fig. 5} shows a cross-coupled transconductance pair with its small-signal model to generate the negative resistance, which is referred to as a "Type $\mathrm{I}$" negative impedance in this paper. The cross-coupled transconductance pair is implemented using two $n$-type transistors, $M_{x}$ and $M_{y}$ (Fig. \ref{fig. 5}(a)), but it can also be implemented using two $p$-type transistors. $g_{m}$ is the transconductance of $M_{x}$ (or $M_{y}$), $V_{x}$ and $V_{y}$ mean their respective gate-source voltages (Fig. \ref{fig. 5}(b)). \footnote{Depending on $V_{x}$ (or $V_{y}$), the current $g_{m}V_{x}$ (or $g_{m}V_{y}$) is generated by $M_{y}$ (or $M_{x}$); the operation principles of a transistor are described in \cite{Razavi2017analog}.} Note that $M_{x}$ and $M_{y}$ are assumed to be designed with the same device size, so their transconductances are equal to $g_{m}$.

The impedance $Z_{t}$ ($=V_{t}/I_{t}$) seen looking into nodes $X$ and $Y$ can be obtained by applying a test voltage $V_{t}$ and observing the resulting current $I_{t}$ in the small-signal model of the cross-coupled transconductance pair (Fig. \ref{fig. 5}(b)). \footnote{To simplify the analysis, the small-signal model is given by ignoring the parasitic capacitances{\textemdash}$C_{gs}$ and $C_{ds}${\textemdash}and assuming that the output resistance $r_{o}$ of $M_{x}$ and $M_{y}$ is infinite.} Applying Kirchhoff's Current Law (KCL) to nodes $X$ and $Y$ in the small-signal model, the two equations are obtained as follows: $I_{t}=g_{m}V_{y}$ and $-I_{t}=g_{m}V_{x}$. The subtraction of the two currents is then given by $2I_{t}=g_{m}(V_{y}-V_{x})$. Since $V_{t}$ is $V_{x}-V_{y}$, the impedance $V_{t}/I_{t}$ is obtained as $-2/g_{m}$. As a result, the impedance of Type $\mathrm{I}$ is expressed as the negative resistance, $-2/g_{m}$. Assuming that $R_{p}>0$ and $R_{n}=-2/g_{m}$ in Fig. \ref{fig. 4}, the effective parallel resistance remains negative as long as $(2/g_{m})<R_{p}$, generating oscillation at $1/(2{\pi}\sqrt{L_{p}C_{p}})$. This means that as $R_{p}$ increases, a higher $g_{m}$ is required to sustain oscillation, which affects the design area, power dissipation, and parasitic capacitance.

\subsection{\color{myblue}Transconductance Modulation}
\begin{figure}[h!]
\centering\includegraphics[scale = 0.55]{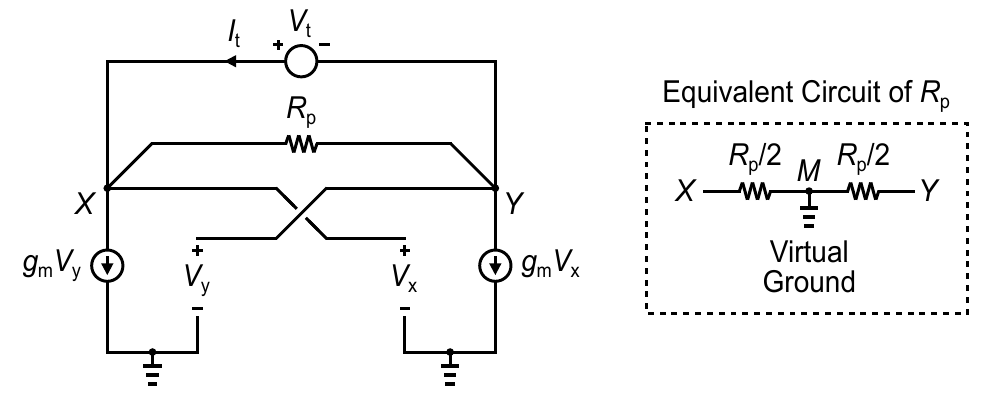}
\caption{{\color{myred}\textbf{Transconductance modulation.}} Small-signal model of Type $\mathrm{I}$ with a parallel resistor.}\label{fig. 6}
\end{figure}

To analyze how the transconductance of Type $\mathrm{I}$ can be affected by a parallel resistor $R_{p}$, the small-signal model of Type $\mathrm{I}$ is modified as shown in Fig. \ref{fig. 6}, where $R_{p}$ is the same as the parallel resistor as depicted in Fig. \ref{fig. 1}. Note that since the Type $\mathrm{I}$ configuration is fully differential, $R_{p}$ can be split into two series resistors, and the midpoint of these two resistors becomes a virtual ground, making the potential at node $M$ equal to zero (Fig. \ref{fig. 6}). This means that if the circuit configuration is fully differential (or symmetric), $R_{p}$ is expressed as the series combination of $R_{p}/2$ and $R_{p}/2$ while maintaining $V_{M}=0$. To find the impedance seen looking into nodes $X$ and $Y$, applying $V_{t}$ to nodes $X$ and $Y$, the KCL result is given by
  
\begin{align}
I_{t}&=g_{m}V_{y}+\frac{V_{x}}{R_{p}/2}\\
-I_{t}&=g_{m}V_{x}+\frac{V_{y}}{R_{p}/2}
\end{align}

By subtracting the two equations above and applying $V_{t}=V_{x}-V_{y}$, the result is $2I_{t}=-g_{m}V_{t}+(2V_{t}/R_{p})$. Therefore, the impedance $Z_{t}$ is given by

\begin{align}
\frac{V_{t}}{I_{t}}&=-\frac{2}{\left(g_{m}-\frac{2}{R_{p}}\right)}\\
&=-\frac{2}{G_{m}}
\end{align}

{\noindent}where $G_{m}$ is expressed as $g_{m}-(2/R_{p})$, which means the effective transconductance considering $R_{p}$. Compared to the Type $\mathrm{I}$ impedance expressed as $-2/g_{m}$, which is the inverse of $g_{m}$ and multiplied by $-2$ (Fig. \ref{fig. 5}), the impedance of Type $\mathrm{I}$ with $R_{p}$ can be expressed as the inverse of the effective transconductance $G_{m}$ multiplied by $-2$ (Fig. \ref{fig. 6}). To maintain negative resistance of $-2/G_{m}$, $G_{m}$ must remain positive, which means that $g_{m}-(2/R_{p})>0$ in Equation (7). Rearranging $g_{m}-(2/R_{p})>0$, the result is given by $(2/g_{m})<R_{p}$, which is the same as the resistance condition to generate oscillation in Fig. \ref{fig. 4}. $G_{m}$ is therefore modulated depending on $R_{p}$, changing the impedance $-2/G_{m}$. In Equation (7), the impedance can be approximated as $-2/g_{m}$ if $R_{p}$ is sufficiently high, where this approximation is equivalent to the impedance of Type $\mathrm{I}$ shown in Fig. \ref{fig. 5}(b).

\subsection{\color{myblue}Oscillation with Type $\mathrm{I}$}
According to the oscillation principle in the previous section using the parallel RLC circuit as depicted in Fig. \ref{fig. 4}, when the resistance conditions{\textemdash}$R_{n}<0$, $R_{p}>0$, and $|R_{n}|<|R_{p}|${\textemdash}are satisfied, the equivalent resistance of $R_{p}$ and $R_{n}$ connected in parallel remains negative, resulting in oscillation at $1/(2{\pi}\sqrt{L_{p}C_{p}})$. Assuming that $R_{n}$ is connected with $R_{p}$ in parallel, where $R_{n}$ is implemented by Type $\mathrm{I}$ shown in Fig. \ref{fig. 5}, the equivalent resistance is obtained by $(-2/g_{m})||R_{p}$. This equivalent resistance can be reduced to $(-2R_{p}/g_{m})/[(-2/g_{m})+R_{p}]$, resulting in $-2/[g_{m}-(2/R_{p})]$, which is equal to $-2/G_{m}$. Recall that $g_{m}-(2/R_{p})$ means the effective transconductance $G_{m}$ (Fig. \ref{fig. 6}).

Therefore, oscillation can be sustained at $1/(2{\pi}\sqrt{L_{p}C_{p}})$ as long as $(2/g_{m})<R_{p}$ (or $|R_{n}|<|R_{p}|$) in a parallel RLC circuit consisting of $R_{n}$ ($=-2/g_{m}$), $R_{p}$, $L_{p}$, and $C_{p}$. In this parallel RLC circuit, $R_{n}$ can be implemented using the active devices, e.g., two $n$-type transistors (Fig. \ref{fig. 5}), whereas $R_{p}$, $L_{p}$, and $C_{p}$ are passive components. Ideally, the oscillation frequency is not affected by $R_{n}$ because oscillation occurs only when the impedance magnitudes of $L_{p}$ and $C_{p}$ are equal each other by ${\omega}L_{p}=({\omega}C_{p})^{-1}$. However, the oscillation amplitude is affected by $R_{n}$ because the equivalent resistance $(-2/g_{m})||R_{p}$ controls energy to induce oscillation. When $|R_{n}|$ ($=2/g_{m}$) is less than $|R_{p}|$, $(-2/g_{m})||R_{p}$ remains negative, and oscillation is therefore sustained. But, as $|R_{n}|$ increases and becomes larger than $|R_{p}|$, $(-2/g_{m})||R_{p}$ changes from negative to positive. This positive resistance dissipates energy that would otherwise be supplied to $L_{p}$ and $C_{p}$, and eventually the oscillation amplitude vanishes. This also means that if the magnitude of $g_{m}$ is not sufficient, $R_{n}$ cannot compensate for $R_{p}$, $(-2/g_{m})||R_{p}$ becomes positive, and eventually the oscillation degrades. However, a sufficiently large $g_{m}$ induces the design area, power dissipation, and parasitic capacitance, so $R_{p}$ must be controlled within an acceptable performance range.

Also, the resonant frequency $1/(2{\pi}\sqrt{L_{p}C_{p}})$ should be appropriately chosen according to an achievable $g_{m}$. This is because $R_{p}$ remains constant over the entire frequency range, whereas $g_{m}$ shows frequency dependence. $R_{p}$ can be compensated by a sufficient $g_{m}$ at low frequencies, maintaining $(-2/g_{m})||R_{p}<0$. But as the operating frequency increases, $g_{m}$ degrades after passing a certain frequency range, turning $(-2/g_{m})||R_{p}$ from negative to positive, and oscillation disappears at high frequencies even if $L_{p}$ and $C_{p}$ are chosen to set a higher resonant frequency. Therefore, the resonant frequency range is limited by the achievable $g_{m}$. Since CMOS technology determines a maximum achievable $g_{m}$, adequate CMOS technology must be used by considering a targeted resonant frequency and $R_{p}$. 
\section{\color{myblue}\Large{N}\large{EGATIVE} \Large{R}\large{EACTANCE}}
In the calculations of the Type $\mathrm{I}$ impedance in Figs. \ref{fig. 5} and \ref{fig. 6}, the impedance is given by the negative resistance, $-2/g_{m}$ or $-2/G_{m}$, without the reactance. \footnote{Impedance $Z$ consists of resistance $R$ and reactance $X$, as $R+jX$. Here, $R$ is independent of frequency and $X$ varies with frequency.} However, the reactance is essential for generating oscillation because oscillation occurs at a frequency where the reactance is zero and the resistance remains negative. This reactance can be implemented using the passive components, $L_{p}$ and $C_{p}$, as depicted in Fig. \ref{fig. 4}. Therefore, the Type $\mathrm{I}$ impedance cannot solely generate oscillation without $L_{p}$ and $C_{p}$.

To generate oscillation while maintaining the negative resistance and incorporating the reactance, let us assume that a negative impedance $Z_{ni}$ is implemented by a series combination of an active resistance $R_{ni}$, an active inductance $L_{ni}$, and an active capacitance $C_{ni}$ without using passive components such as $R_{p}$, $L_{p}$, and $C_{p}$. This negative impedance is then expressed as follows:

\begin{align}
Z_{ni}&=R_{ni}+j{\omega}L_{ni}+\frac{1}{j{\omega}C_{ni}}\\
&=R_{ni}+j\underbrace{\left(\frac{{\omega}^2L_{ni}C_{ni}-1}{{\omega}C_{ni}}\right)}_{X_{ni}}
\end{align}

{\noindent}where $R_{ni}$, $L_{ni}$, and $C_{ni}$ remain negative, and accordingly, the reactance $X_{ni}$ also remains negative. \footnote{Since $R_{ni}$, $L_{ni}$, and $C_{ni}$ will be implemented using transconductive devices later, the term "active" is used to describe these components.} Compared to the Type $\mathrm{I}$ impedance, which consists only of the resistance $-2/g_{m}$ or $-2/G_{m}$, $Z_{ni}$ is composed of the resistance $R_{ni}$ and the reactance $X_{ni}$, allowing oscillation to occur even in the absence of passive components. Then, in this paper, $Z_{ni}${\textemdash}a series combination of $R_{ni}$, $L_{ni}$, and $C_{ni}${\textemdash}will be referred to as a Type $\mathrm{II}$ impedance.

According to the series resonance in Section $\mathrm{I}$, $Z_{ni}$ can be analyzed using voltage phasors of $L_{ni}$ and $C_{ni}$. Since $L_{ni}$ and $C_{ni}$ share a same current, the voltage phasors of $L_{ni}$ and $C_{ni}$ are expressed as $j{\omega}L_{ni}$ and $-j({\omega}C_{ni})^{-1}$, respectively, showing a $180^{\circ}$ phase difference as depicted in Fig. \ref{fig. 2}(a). Note that in the parallel RLC circuit (Fig. \ref{fig. 2}(b)), since the voltage drops across each component are equal, the current phasors of $L_{p}$ and $C_{p}$ are $-j({\omega}L_{p})^{-1}$ and $j{\omega}C_{p}$, respectively, exhibiting a phase difference of $180^{\circ}$ in the phasor diagram drawn in the current domain (Fig. \ref{fig. 2}(c)).

\begin{figure}[h!]
\centering\includegraphics[scale = 0.55]{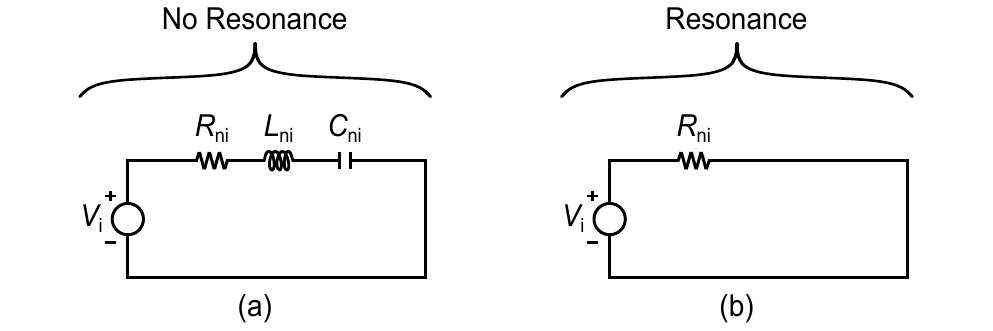}
\caption{{\color{myred}\textbf{$Z_{ni}$ with a voltage impulse $V_{i}$.}} (a) $Z_{ni}$ under no resonance. (b) $Z_{ni}$ under resonance.}\label{fig. 7}
\end{figure}

Assuming that a voltage impulse $V_{i}$ is applied to $Z_{ni}$ as shown in Fig. \ref{fig. 7}(a), the voltage drop across the $L_{ni}$ and $C_{ni}$ is minimized at a frequency $f_{ni}$, where the magnitudes of the two voltage phasors are equal and cancel each other. \footnote{In the Type $\mathrm{II}$ impedance, $f_{ni}$ is calculated as $1/(2{\pi}\sqrt{L_{ni}C_{ni}})$ using the equations, ${\omega}_{ni}L_{ni}=({\omega}_{ni}C_{ni})^{-1}$ and ${\omega}_{ni}=2{\pi}f_{ni}$.} This means that the sum of the two impedances of $L_{ni}$ and $C_{ni}$ becomes negligible at $f_{ni}$ by the negligible voltage drop across the $L_{ni}$ and $C_{ni}$, which can therefore be expressed as depicted in Fig. \ref{fig. 7}(b). This minimum impedance can also be obtained from Equation (10) if ${\omega}$ is $1/\sqrt{L_{ni}C_{ni}}$.

Consequently, the minimum impedance of the $L_{ni}$ and $C_{ni}$ means that the maximum current $i_{ni,max}$ must be drawn through $Z_{ni}$ at $f_{ni}$, which can be described as shown in Fig. \ref{fig. 7}(b). In other words, $R_{ni}$ draws $i_{ni,max}$ under resonance, achieving the maximum power $i_{ni,max}^2R_{ni}$ at $f_{ni}$. Unlike a positive (called "passive") resistance, which dissipates energy, a negative (called "active") resistance supplies energy \cite{Chua1987resistor}. Applying this resistance principle and assuming that $R_{ni}$ remains negative and ${\omega}_{ni}L_{ni}=({\omega}_{ni}C_{ni})^{-1}$, the maximum power $i_{ni,max}^2R_{ni}$ at $f_{ni}$ is sustained within $Z_{ni}$ rather than vanishing, leading to oscillation at $f_{ni}$.

\begin{figure}[h!]
\centering\includegraphics[scale = 0.55]{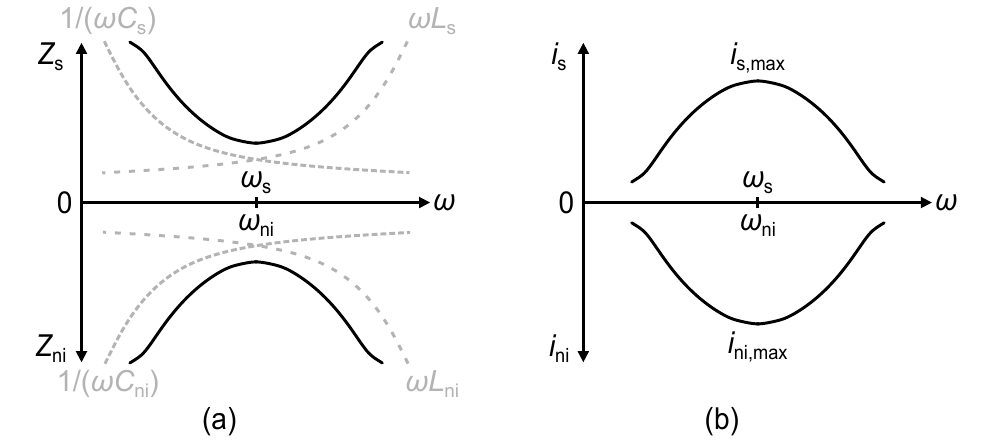}
\caption{{\color{myred}\textbf{Positive series impedance $Z_{s}$ and negative series impedance $Z_{ni}$.}} (a) Comparison of impedance magnitudes. (b) Comparison of ac current magnitudes.}\label{fig. 8}
\end{figure}

To clarify the concept of the negative series impedance $Z_{ni}$, the magnitudes of impedance and ac current are compared by using $Z_{ni}$ and the positive series impedance $Z_{s}$ (Fig. \ref{fig. 8}). Recall that $Z_{s}$ is the series RLC circuit expressed as $R_{s}+j{\omega}L_{s}+(j{\omega}C_{s})^{-1}$ as discussed in Section $\mathrm{I}$. Considering that an impedance magnitude is determined by the largest value in a series circuit, the magnitude of $Z_{ni}$ has a symmetrical profile with respect to the magnitude of $Z_{s}$ (Fig. \ref{fig. 8}(a)). ${\omega}_{s}$ and ${\omega}_{ni}$ are the resonant frequencies of $Z_{s}$ and $Z_{ni}$, respectively. $i_{s}$ and $i_{ni}$ mean the ac currents flowing through $Z_{s}$ and $Z_{ni}$, respectively. The magnitudes of $Z_{s}$ and $Z_{ni}$ are minimized at ${\omega}_{s}$ and ${\omega}_{ni}$, respectively, meaning that the maximum ac currents of $Z_{s}$ and $Z_{ni}$ are achieved as $i_{s,max}$ and $i_{ni,max}$ at ${\omega}_{s}$ and ${\omega}_{ni}$, respectively (Fig. \ref{fig. 8}(b)). As depicted in Fig. \ref{fig. 8}(b), $i_{s,max}$ is the ac current maximized in the positive direction ($i_{s}>0$ and $i_{s,max}>0$), whereas $i_{ni,max}$ is the ac current maximized in the negative direction ($i_{ni}<0$ and $i_{ni,max}<0$). From Equation (10) and Fig. \ref{fig. 8}(b), the power expressed as $i_{ni}^2R_{ni}$ is maximized in the negative direction at ${\omega}_{ni}$, achieving $i_{ni,max}^2R_{ni}$. Since $i_{ni,max}^2R_{ni}<0$ by $R_{ni}<0$, $i_{ni,max}^2R_{ni}$ is sustained rather than dissipated through $R_{ni}$, leading to oscillation at ${\omega}_{ni}$ among the entire frequency range.

When analyzing the oscillation phenomena using Figs. \ref{fig. 4} and \ref{fig. 7}, the current impulse $I_{i}$ is applied to the parallel RLC circuit, and the voltage impulse $V_{i}$ is applied to the series RLC circuit. This is because the parallel RLC circuit shows the phase difference in the current domain, whereas the series RLC circuit shows the phase difference in the voltage domain. Therefore, $I_{i}$ and $V_{i}$ are employed for the parallel and series RLC circuits, respectively, to understand the oscillation phenomena by considering the phase shift in each circuit model. 
\subsection{\color{myblue}Negative Impedance{\textemdash}Type $\mathrm{II}$}
\begin{figure}[h!]
\centering\includegraphics[scale = 0.55]{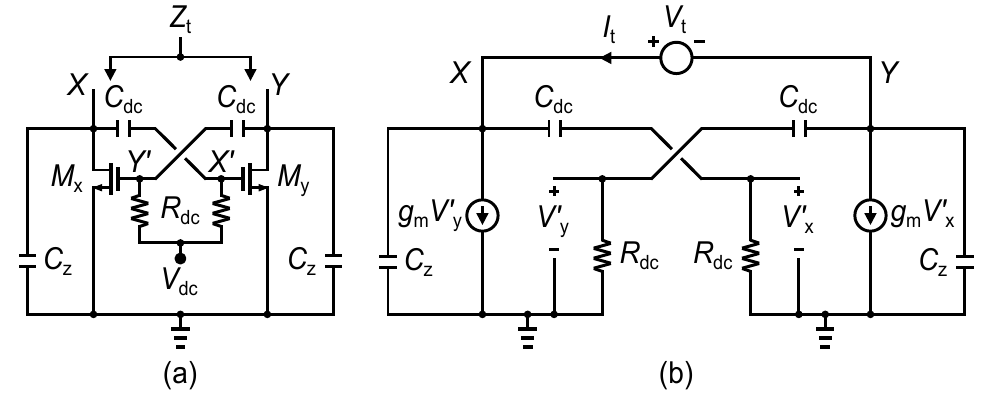}
\caption{{\color{myred}\textbf{Negative impedance{\textemdash}Type $\mathrm{II}$.}} (a) Implementation of Type $\mathrm{II}$. (b) Its small-signal model.}\label{fig. 9}
\end{figure}

To implement the Type $\mathrm{II}$ impedance, $Z_{ni}$, the Type $\mathrm{I}$ configuration is modified as shown in Fig. \ref{fig. 9}(a). Compared to the Type $\mathrm{I}$ configuration (Fig. \ref{fig. 5}(a)), to implement $L_{ni}$ and $C_{ni}$ in Equation (9), the decoupling capacitor $C_{dc}$, dc bias resistor $R_{dc}$, and capacitor $C_{z}$ are integrated around the cross-coupled transconductance pair consisting of $M_{x}$ and $M_{y}$ (Fig. \ref{fig. 9}(a)).

To obtain the Type $\mathrm{II}$ impedance seen looking into nodes $X$ and $Y$, small-signal analysis is conducted by applying a test voltage $V_{t}$ and observing the resulting current $I_{t}$ as shown in Fig. \ref{fig. 9}(b). Applying $V_{t}$, two currents are given by

\begin{align}
I_{t}&=g_{m}V_{y}^{\prime}+sC_{z}V_{x}+\frac{V_{x}}{Z_{dc}}\\
-I_{t}&=g_{m}V_{x}^{\prime}+sC_{z}V_{y}+\frac{V_{y}}{Z_{dc}}
\end{align}

{\noindent}where $V_{x}^{\prime}$ is the gate voltage of $M_{y}$, $V_{y}^{\prime}$ is the gate voltage of $M_{x}$, and $Z_{dc}$ is the series impedance of $R_{dc}$ and $C_{dc}$. Using $V_{t}=V_{x}-V_{y}$, $V_{x}^{\prime}=V_{x}[sR_{dc}C_{dc}/(1+sR_{dc}C_{dc})]$, $V_{y}^{\prime}=V_{y}[sR_{dc}C_{dc}/(1+sR_{dc}C_{dc})]$, and $Z_{dc}=(1+sR_{dc}C_{dc})/sC_{dc}$, the above two currents can be subtracted as follows:

\begin{align}
2I_{t}=-g_{m}V_{t}\frac{sR_{dc}C_{dc}}{1+sR_{dc}C_{dc}}+sC_{z}V_{t}+V_{t}\frac{sC_{dc}}{1+sR_{dc}C_{dc}}
\end{align}

{\noindent}Rearranging the above equation, $V_{t}/I_{t}$ can be given by

\begin{align}
\frac{V_{t}}{I_{t}}=\frac{2(1+sR_{dc}C_{dc})}{-g_{m}sR_{dc}C_{dc}+sC_{z}(1+sR_{dc}C_{dc})+sC_{dc}}
\end{align}

{\noindent}Using $s=j{\omega}$ and $j^{2}=-1$, Equation (14) is modified as

\begin{align}
\frac{2(1+j{\omega}R_{dc}C_{dc})}{-{\omega}^2R_{dc}C_{dc}C_{z}+j{\omega}(-g_{m}R_{dc}C_{dc}+C_{z}+C_{dc})}
\end{align}

Multiplying the complex conjugate of the denominator of Equation (15) by the numerator and denominator, $V_{t}/I_{t}$ can be expressed as the resistance $R_{ni}$ and reactance $X_{ni}$:

\begin{align}
\frac{V_{t}}{I_{t}}&=R_{ni}+jX_{ni}\\
R_{ni}&=\frac{-2{\omega}^2R_{dc}C_{dc}C_{z}+2{\omega}^2R_{dc}C_{dc}K_{ni}}{{\omega}^4(R_{dc}C_{dc}C_{z})^2+{\omega}^2K_{ni}^2}\\
X_{ni}&=\frac{-2{\omega}^3R_{dc}^2C_{dc}^2C_{z}-2{\omega}K_{ni}}{{\omega}^4(R_{dc}C_{dc}C_{z})^2+{\omega}^2K_{ni}^2}
\end{align}

{\noindent}where $K_{ni}=-g_{m}R_{dc}C_{dc}+C_{z}+C_{dc}$. By choosing $g_{m}R_{dc}C_{dc}$ to be sufficiently larger than $C_{z}+C_{dc}$, $K_{ni}$ can be approximated as $-g_{m}R_{dc}C_{dc}$. Then, by using $K_{ni}{\,}{\approx}{\,}-g_{m}R_{dc}C_{dc}$ and choosing $[1/({\omega}C_{z})]>(1/g_{m})$, the approximations of $R_{ni}$ and $X_{ni}$ can be given by

\begin{align}
R_{ni}&{\,}{\approx}{\,}\frac{-2{\omega}^2R_{dc}C_{dc}C_{z}-2{\omega}^2g_{m}R_{dc}^2C_{dc}^2}{{\omega}^4(R_{dc}C_{dc}C_{z})^2+{\omega}^2(g_{m}R_{dc}C_{dc})^2}\\
&{\,}{\approx}{\,}\frac{-2R_{dc}C_{dc}C_{z}-2g_{m}R_{dc}^2C_{dc}^2}{{\omega}^2(R_{dc}C_{dc}C_{z})^2+(g_{m}R_{dc}C_{dc})^2}\\
&{\,}{\approx}{\,}\frac{-2R_{dc}C_{dc}C_{z}-2g_{m}R_{dc}^2C_{dc}^2}{(g_{m}R_{dc}C_{dc})^2}\\
X_{ni}&{\,}{\approx}{\,}\frac{-2{\omega}^3R_{dc}^2C_{dc}^2C_{z}+2{\omega}g_{m}R_{dc}C_{dc}}{{\omega}^4(R_{dc}C_{dc}C_{z})^2+{\omega}^2(g_{m}R_{dc}C_{dc})^2}\\
&{\,}{\approx}{\,}\frac{-2{\omega}^3R_{dc}^2C_{dc}^2C_{z}+2{\omega}g_{m}R_{dc}C_{dc}}{{\omega}^2(g_{m}R_{dc}C_{dc})^2}\\
&=\frac{-2{\omega}C_{z}}{g_{m}^2}+\frac{2}{{\omega}g_{m}R_{dc}C_{dc}}
\end{align}

Recall that the conditions, $g_{m}R_{dc}C_{dc}>(C_{z}+C_{dc})$ and $[1/({\omega}C_{z})]>(1/g_{m})$, the numerator of Equation (21) can be further approximated as $-2g_{m}R_{dc}^2C_{dc}^2$. $R_{ni}$ is therefore approximated as $-2/g_{m}$ from Equation (21). As a result of the above approximations, $R_{ni}+jX_{ni}$ from Equation (16) is modified to

\begin{align}
\underbrace{-\frac{2}{g_{m}}}_{R_{ni}}+j\underbrace{\left(\frac{-2{\omega}C_{z}}{g_{m}^2}+\frac{2}{{\omega}g_{m}R_{dc}C_{dc}}\right)}_{X_{ni}}
\end{align}

{\noindent}Rearranging Equation (25) and using $j^{-1}=-j$, $X_{ni}$ can be decomposed into $L_{ni}$ and $C_{ni}$ as follows:

\begin{align}
\underbrace{-\frac{2}{g_{m}}}_{R_{ni}}+j{\omega}\underbrace{\frac{-2C_{z}}{g_{m}^2}}_{L_{ni}}+\frac{1}{j{\omega}\underbrace{\frac{-g_{m}R_{dc}C_{dc}}{2}}_{C_{ni}}}
\end{align}

{\noindent}From Equation (26), the active components are given by $R_{ni}=-2/g_{m}$, $L_{ni}=-2C_{z}/(g_{m}^2)$, and $C_{ni}=-g_{m}R_{dc}C_{dc}/2$. All components are negative and can be expressed in the form of Equation (9). Therefore, according to Equation (10) and Fig. \ref{fig. 8}, the Type $\mathrm{II}$ configuration shown in Fig. \ref{fig. 9} generates oscillation at $f_{ni}$ as follows:

\begin{align}
f_{ni}&=\frac{1}{2{\pi}\sqrt{L_{ni}C_{ni}}}\\
&=\frac{1}{2{\pi}}\sqrt{\frac{g_{m}}{R_{dc}C_{dc}C_{z}}}
\end{align}

{\noindent}From Equation (28), the oscillation frequency is proportional to $g_{m}$, while the components of $R_{dc}$, $C_{dc}$, and $C_{z}$ decrease the frequency. Compared to the Type $\mathrm{I}$ configuration shown in Fig. \ref{fig. 5}, the Type $\mathrm{II}$ configuration shown in Fig. \ref{fig. 9} not only provides a stable dc bias setting by $R_{dc}$ and $C_{dc}$, but can also perform self-oscillation without dependent on passive components such as $R_{p}$, $L_{p}$, and $C_{p}$ shown in Fig. \ref{fig. 4}.

To simplify the calculation to obtain $f_{ni}$ from Equation (14), the numerator of $X_{ni}$ from Equation (22) can be used by finding ${\omega}_{ni}$ where the numerator becomes zero. Recall that the oscillation arising from the Type $\mathrm{II}$ configuration is induced under the condition that the reactance $X_{ni}$ from Equation (10) becomes zero. Accordingly, by setting the numerator of Equation (22) to zero, ${\omega}_{ni}$ and $f_{ni}$ can be obtained through the following calculation:

\begin{align}
-2{\omega}^3R_{dc}^2C_{dc}^2C_{z}+2{\omega}g_{m}R_{dc}C_{dc}=0
\end{align}

{\noindent}Rearranging Equation (29) yields

\begin{align}
g_{m}={\omega}^2R_{dc}C_{dc}C_{z}
\end{align}

\begin{align}
{\omega}_{ni}=\sqrt{\frac{g_{m}}{R_{dc}C_{dc}C_{z}}}{\hspace{7pt}\text{and}\hspace{7pt}}f_{ni}=\frac{1}{2{\pi}}\sqrt{\frac{g_{m}}{R_{dc}C_{dc}C_{z}}}
\end{align} 
\subsection{\color{myblue}Analysis including a Parallel Resistor}
\begin{figure}[h!]
\centering\includegraphics[scale = 0.55]{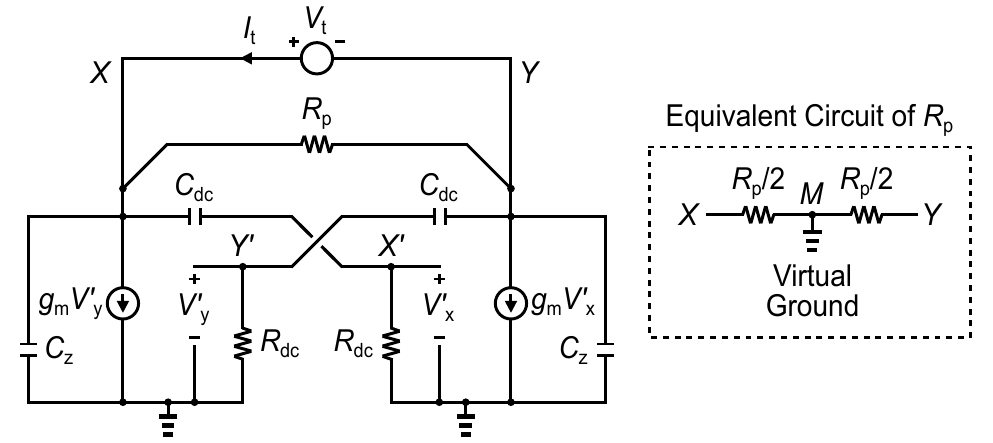}
\caption{{\color{myred}\textbf{Type $\mathrm{II}$ with a parallel resistor.}} Small-signal model of Type $\mathrm{II}$ and an equivalent circuit of $R_{p}$.}\label{fig. 10}
\end{figure}

To understand how the parallel resistor $R_{p}$ affects the frequency operation of Type $\mathrm{II}$, an impedance analysis is conducted using the small-signal model of Type $\mathrm{II}$ and $R_{p}$ (Fig. \ref{fig. 10}). Since the configuration of Type $\mathrm{II}$ is fully differential, $R_{p}$ can be decomposed into two series resistors using $R_{p}/2$, and the midpoint, $M$, of the two resistors becomes a virtual ground ($V_{M}=0$).

Applying KCL with a test voltage $V_{t}$ to nodes $X$ and $Y$, the following two currents are given by

\begin{align}
I_{t}&=g_{m}V_{y}^{\prime}+sC_{z}V_{x}+\frac{V_{x}}{Z_{dc}}+\frac{V_{x}}{R_{p}/2}\\
-I_{t}&=g_{m}V_{x}^{\prime}+sC_{z}V_{y}+\frac{V_{y}}{Z_{dc}}+\frac{V_{y}}{R_{p}/2}
\end{align}

After subtracting the above two currents and applying $V_{t}=V_{x}-V_{y}$, $V_{x}^{\prime}=V_{x}[sR_{dc}C_{dc}/(1+sR_{dc}C_{dc})]$, $V_{y}^{\prime}=V_{y}[sR_{dc}C_{dc}/(1+sR_{dc}C_{dc})]$, and $Z_{dc}=(1+sR_{dc}C_{dc})/sC_{dc}$, the impedance is given by

\begin{align}
\frac{V_{t}}{I_{t}}&=\frac{2R_{p}(1+sR_{dc}C_{dc})}{\left[\begin{array}{c}
-g_{m}R_{p}sR_{dc}C_{dc}+sR_{p}C_{z}(1+sR_{dc}C_{dc})\\
+sR_{p}C_{dc}+2(1+sR_{dc}C_{dc})
\end{array}\right]}\\
&=\frac{2R_{p}(1+j{\omega}R_{dc}C_{dc})}{\left[\begin{array}{c}
(2-{\omega}^2R_{dc}C_{dc}R_{p}C_{z})-j{\omega}g_{m}R_{p}R_{dc}C_{dc}\\
+j{\omega}(R_{p}C_{z}+R_{p}C_{dc}+2R_{dc}C_{dc})
\end{array}\right]}\\
&{\,}{\approx}{\,}\frac{2R_{p}(1+j{\omega}R_{dc}C_{dc})}{(2-{\omega}^2R_{dc}C_{dc}R_{p}C_{z})-j{\omega}g_{m}R_{p}R_{dc}C_{dc}}
\end{align}

{\noindent}By choosing $g_{m}R_{p}R_{dc}C_{dc}>(R_{p}C_{z}+R_{p}C_{dc}+2R_{dc}C_{dc})$, the denominator of Equation (35) is approximated by that of Equation (36). Then, multiplying the complex conjugate of the denominator of Equation (36) by the numerator of Equation (36), the numerator of $V_{t}/I_{t}$ is given by

\begin{align}
\begin{array}{c}
2R_{p}(1+j{\omega}R_{dc}C_{dc})\\
{\times}\left[(2-{\omega}^2R_{dc}C_{dc}R_{p}C_{z})+j{\omega}g_{m}R_{p}R_{dc}C_{dc}\right]
\end{array}
\end{align}

{\noindent}The imaginary part of Equation (37){\textemdash}the numerator of the reactance of $V_{t}/I_{t}${\textemdash}is obtained as

\begin{align}
2R_{p}\left[{\omega}R_{dc}C_{dc}(2-{\omega}^2R_{dc}C_{dc}R_{p}C_{z})+{\omega}g_{m}R_{p}R_{dc}C_{dc}\right]
\end{align}

Equating the above imaginary part to zero means that the reactance of Equation (36) becomes zero, which can be used to find ${\omega}_{ni}$. Recall that the oscillation frequency of Type $\mathrm{II}$ shown in Fig. \ref{fig. 9} can be obtained where $X_{ni}=0$ in Equation (22). Setting Equation (38) to zero is given by

\begin{align}
{\omega}R_{dc}C_{dc}(2-{\omega}^2R_{dc}C_{dc}R_{p}C_{z})+{\omega}g_{m}R_{p}R_{dc}C_{dc}&=0\\
\frac{2}{R_{p}}-{\omega}^2R_{dc}C_{dc}C_{z}+g_{m}&=0
\end{align}

{\noindent}Solving Equation (40), ${\omega}_{ni}$ and $f_{ni}$ are obtained as follows:

\begin{align}
{\omega}_{ni}=\sqrt{\frac{g_{m}+\frac{2}{R_{p}}}{R_{dc}C_{dc}C_{z}}}{\hspace{7pt}\text{and}\hspace{7pt}}f_{ni}=\frac{1}{2{\pi}}\sqrt{\frac{g_{m}+\frac{2}{R_{p}}}{R_{dc}C_{dc}C_{z}}}
\end{align}

{\noindent}Compared to $f_{ni}$ derived without $R_{p}$ in Equation (28), $f_{ni}$ in Equation (41) is obtained by adding $2/R_{p}$ to $g_{m}$. Assuming that $R_{p}$ is sufficiently large, $f_{ni}$ in Equation (41) is approximated as $f_{ni}$ in Equation (28). On the other hand, as $R_{p}$ decreases, $f_{ni}$ in Equation (41) tends to increase. However, a continuous decrease of $R_{p}$ makes a short circuit between nodes $X$ and $Y$ (Fig. \ref{fig. 10}), thereby causing the circuit to stop working. Therefore, $R_{p}$ can be employed to control $f_{ni}$, but the tuning range of $R_{p}$ is limited because a low value of $R_{p}$ induces a short circuit and destroys the entire circuit operation.

\subsection{\color{myblue}Analysis including a Parallel Inductor}
\begin{figure}[h!]
\centering\includegraphics[scale = 0.55]{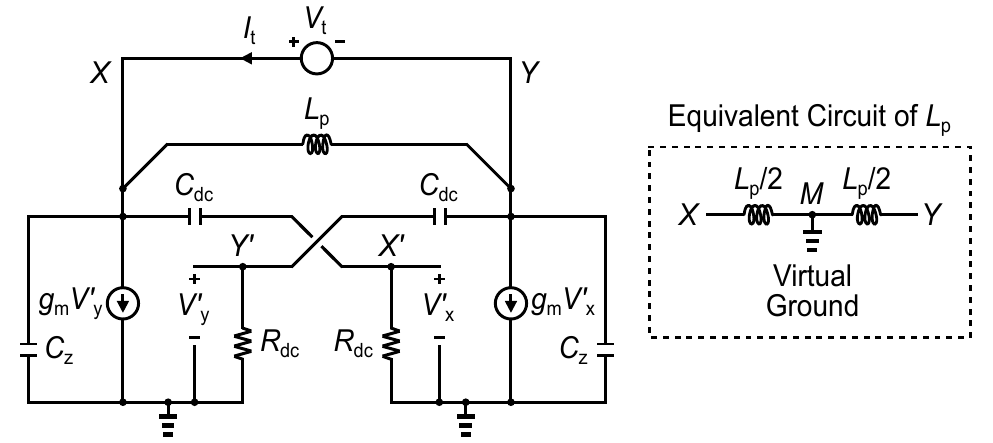}
\caption{{\color{myred}\textbf{Type $\mathrm{II}$ with a parallel inductor.}} Small-signal model of Type $\mathrm{II}$ and an equivalent circuit of $L_{p}$.}\label{fig. 11}
\end{figure}

To understand how the parallel inductor $L_{p}$ affects $f_{ni}$, an impedance analysis is conducted using the small-signal model of Type $\mathrm{II}$ and $L_{p}$ (Fig. \ref{fig. 11}). Similar to the small-signal model of Type $\mathrm{II}$ with $R_{p}$, $L_{p}$ can be decomposed into two series inductors using $L_{p}/2$, and the midpoint, $M$, of the two series inductors is a virtual ground ($V_{M}=0$).

Applying KCL with a test voltage $V_{t}$ to nodes $X$ and $Y$, the following two currents are obtained as

\begin{align}
I_{t}&=g_{m}V_{y}^{\prime}+sC_{z}V_{x}+\frac{V_{x}}{Z_{dc}}+\frac{V_{x}}{sL_{p}/2}\\
-I_{t}&=g_{m}V_{x}^{\prime}+sC_{z}V_{y}+\frac{V_{y}}{Z_{dc}}+\frac{V_{y}}{sL_{p}/2}
\end{align}

Similar to the calculation for Equation (34) derived using the equations of $V_{t}$, $V_{x}^{\prime}$, $V_{y}^{\prime}$, and $Z_{dc}$, the above two currents are subtracted to obtain $V_{t}/I_{t}$ as follows:

\begin{align}
\frac{V_{t}}{I_{t}}&=\frac{2sL_{p}(1+sR_{dc}C_{dc})}{\left[\begin{array}{c}
-s^2g_{m}R_{dc}C_{dc}L_{p}+s^2L_{p}C_{z}(1+sR_{dc}C_{dc})\\
+s^2L_{p}C_{dc}+2(1+sR_{dc}C_{dc})
\end{array}\right]}\\
&=\frac{2(-{\omega}^2R_{dc}C_{dc}L_{p}+j{\omega}L_{p})}{\left[\begin{array}{c}
{\omega}^2g_{m}R_{dc}C_{dc}L_{p}-{\omega}^2L_{p}C_{z}-{\omega}^2L_{p}C_{dc}+2\\
-j({\omega}^3R_{dc}C_{dc}L_{p}C_{z}-{\omega}2R_{dc}C_{dc})
\end{array}\right]}
\end{align}

{\noindent}By choosing $g_{m}R_{dc}C_{dc}L_{p}>(L_{p}C_{z}+L_{p}C_{dc})$ in the denominator of Equation (45), $V_{t}/I_{t}$ can be approximated as

\begin{align}
\frac{2(-{\omega}^2R_{dc}C_{dc}L_{p}+j{\omega}L_{p})}{\left[\begin{array}{c}
{\omega}^2g_{m}R_{dc}C_{dc}L_{p}+2\\
-j({\omega}^3R_{dc}C_{dc}L_{p}C_{z}-{\omega}2R_{dc}C_{dc})
\end{array}\right]}
\end{align}

Similar to the calculations in finding ${\omega}_{ni}$ in the previous sections, the oscillation frequency influenced by $L_{p}$ can be obtained by setting the reactance $X_{ni}$ in Equation (46) to zero. Then, from Equation (46), multiplying the complex conjugate of the denominator by the numerator, $X_{ni}$ is

\begin{align}
\frac{\left[\begin{array}{c}
{\omega}L_{p}({\omega}^2g_{m}R_{dc}C_{dc}L_{p}+2)\\
-{\omega}^2R_{dc}C_{dc}L_{p}({\omega}^3R_{dc}C_{dc}L_{p}C_{z}-{\omega}2R_{dc}C_{dc})
\end{array}\right]}{\left[\begin{array}{c}
({\omega}^2g_{m}R_{dc}C_{dc}L_{p}+2)^2\\
+({\omega}^3R_{dc}C_{dc}L_{p}C_{z}-{\omega}2R_{dc}C_{dc})^2
\end{array}\right]}
\end{align}

{\noindent}From Equation (47), the numerator can be rearranged into ${\omega}^3L_{p}X_{ni,num}$ by factoring out ${\omega}^3L_{p}$ as a common factor, and $X_{ni,num}$ is given by

\begin{align}
g_{m}R_{dc}C_{dc}L_{p}+\frac{2}{{\omega}^2}-{\omega}^2R_{dc}^2C_{dc}^2L_{p}C_{z}+2R_{dc}^2C_{dc}^2
\end{align}

{\noindent}Considering the orders of $g_{m}$, $R_{dc}$, $C_{dc}$, $L_{p}$, $C_{z}$, and ${\omega}$, the second term of $X_{ni,num}$ can be negligible. \footnote{For example, $2/({\omega}^2)$ of Equation (48) can be negligible by setting the values as follows: $g_{m}=10{\,}\mathrm{mS}$, $R_{dc}=1{\,}\mathrm{k}{\Omega}$, $C_{dc}=500{\,}\mathrm{fF}$, $L_{p}=15{\,}{\mu}\mathrm{H}$, $C_{z}=200{\,}\mathrm{fF}$, and ${\omega}=2{\pi}(10^{10})$.} Then, from Equation (48), setting $X_{ni,num}=0$, neglecting $2/({\omega}^2)$, and canceling out $R_{dc}C_{dc}$, the equation can be approximated as

\begin{align}
g_{m}L_{p}-{\omega}^2R_{dc}C_{dc}L_{p}C_{z}+2R_{dc}C_{dc}=0
\end{align}

{\noindent}From Equation (49), ${\omega}_{ni}$ and $f_{ni}$ are therefore obtained as

\begin{align}
{\omega}_{ni}=\sqrt{\frac{g_{m}+\frac{2R_{dc}C_{dc}}{L_{p}}}{R_{dc}C_{dc}C_{z}}}{\hspace{2pt}\text{and}\hspace{2pt}}f_{ni}=\frac{1}{2{\pi}}\sqrt{\frac{g_{m}+\frac{2R_{dc}C_{dc}}{L_{p}}}{R_{dc}C_{dc}C_{z}}}
\end{align}

{\noindent}$f_{ni}$ is inversely proportional to $L_{p}$, and the equation of $f_{ni}$ in Equation (50) can be rearranged as

\begin{align}
f_{ni}=\frac{1}{2{\pi}}\sqrt{\frac{g_{m}}{R_{dc}C_{dc}C_{z}}+\frac{2}{L_{p}C_{z}}}
\end{align}

{\noindent}Assuming that $L_{p}$ is sufficiently large, the term $2/(L_{p}C_{z})$ in Equation (51) can be neglected, which is the same as $f_{ni}$ in Equation (28) derived without a parallel component. From the perspective of circuit configuration, as $L_{p}$ increases, its impedance $j{\omega}L_{p}$ increases, eventually resulting in an open circuit when $L_{p}$ becomes sufficiently large. Consequently, the small-signal model of the Type $\mathrm{II}$ configuration including $L_{p}$ can be approximated to the form depicted in Fig. \ref{fig. 9}(b) when $L_{p}$ is sufficiently large.

\subsection{\color{myblue}Analysis including a Parallel Capacitor}
\begin{figure}[h!]
\centering\includegraphics[scale = 0.55]{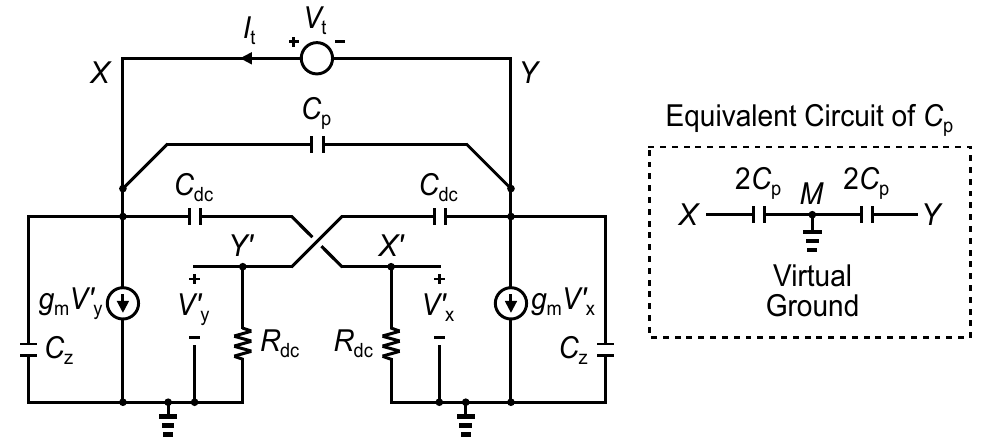}
\caption{{\color{myred}\textbf{Type $\mathrm{II}$ with a parallel capacitor.}} Small-signal model of Type $\mathrm{II}$ and an equivalent circuit of $C_{p}$.}\label{fig. 12}
\end{figure}

Similar to the calculations conducted to obtain $f_{ni}$ with $R_{p}$ or $L_{p}$, to understand the effect of the parallel capacitor $C_{p}$ on the oscillation frequency, an impedance analysis is conducted using the small-signal model of Type $\mathrm{II}$ with $C_{p}$ as shown in Fig. \ref{fig. 12}.

Since the Type $\mathrm{II}$ configuration is fully differential, $C_{p}$ can be decomposed into two series capacitors using $2C_{p}$ with the midpoint, $M$, expressed as a virtual ground. Then, applying KCL with a test voltage $V_{t}$ to nodes $X$ and $Y$ yields the following currents:

\begin{align}
I_{t}&=g_{m}V_{y}^{\prime}+sC_{z}V_{x}+\frac{V_{x}}{Z_{dc}}+s2C_{p}V_{x}\\
-I_{t}&=g_{m}V_{x}^{\prime}+sC_{z}V_{y}+\frac{V_{y}}{Z_{dc}}+s2C_{p}V_{y}
\end{align}

{\noindent}From Equations (52) and (53), since $s2C_{p}$ can merge with $sC_{z}$ by the common factors $V_{x}$ and $V_{y}$, the calculation to obtain $V_{t}/I_{t}$ can be simplified by adding $s2C_{p}$ to $sC_{z}$ in the $V_{t}/I_{t}$ equation derived from the Type $\mathrm{II}$ configuration without a parallel component. Therefore, by utilizing the equations for $V_{t}$, $V_{x}^{\prime}$, $V_{y}^{\prime}$, and $Z_{dc}$, and subtracting the above two currents of Equations (52) and (53), $V_{t}/I_{t}$ for Fig. \ref{fig. 12} can be obtained as follows:

\begin{align}
\frac{2(1+sR_{dc}C_{dc})}{-g_{m}sR_{dc}C_{dc}+s(C_{z}+2C_{p})(1+sR_{dc}C_{dc})+sC_{dc}}
\end{align}

{\noindent}Also, Equation (54) can be directly obtained by adding $s2C_{p}$ to $sC_{z}$ in Equation (14), which is derived from the Type $\mathrm{II}$ configuration without a parallel component. From Equation (54), the remaining calculation steps for finding $f_{ni}$ involve decomposing $V_{t}/I_{t}$ into the resistance $R_{ni}$ and reactance $X_{ni}$, and solving $X_{ni}=0$ to obtain ${\omega}_{ni}$ and $f_{ni}$.

Another approach for finding $f_{ni}$ from Equation (54) is to derive the active inductance $L_{ni}$ and the active capacitance $C_{ni}$ from $X_{ni}$, then $f_{ni}$ is obtained as $1/(2{\pi}\sqrt{L_{ni}C_{ni}})$.

Therefore, utilizing the approximations used to obtain Equations (25) and (29) in Section $\mathrm{III}.A$, and substituting $s(C_{z}+2C_{p})$ for $sC_{z}$, the active inductance $L_{ni}$, active capacitance $C_{ni}$, and oscillation frequency $f_{ni}$ are derived from Equation (54) as follows:

\begin{align}
L_{ni}&=-\frac{2(C_{z}+2C_{p})}{g_{m}^2}\\
C_{ni}&=-\frac{g_{m}R_{dc}C_{dc}}{2}\\
f_{ni}&=\frac{1}{2{\pi}}\sqrt{\frac{g_{m}}{R_{dc}C_{dc}(C_{z}+2C_{p})}}
\end{align}

As depicted in Fig. \ref{fig. 12}, $C_{p}$ connected with Type $\mathrm{II}$ in parallel controls $L_{ni}$ but does not affect $C_{ni}$. Consequently, $f_{ni}$ is inversely proportional to $C_{p}$.

\subsection{\color{myblue}Summary of Type $\mathrm{II}$}
In this section, the small-signal analysis of Type $\mathrm{II}$ is conducted according to four cases: (1) without a parallel component, (2) with $R_{p}$, (3) with $L_{p}$, and (4) with $C_{p}$. $f_{ni}$ for each case is summarized in Table \ref{table:fni}.

\begin{table}[h!]
\caption{Resonant Frequency Summary}\vspace{-10pt}
\centering\begin{tabular}{p{20mm} | p{50mm}}\hline\hline
\addstackgap[7pt]{(a) $f_{ni}$}&$=\frac{1}{2{\pi}}\sqrt{\frac{g_{m}}{R_{dc}C_{dc}C_{z}}}$\\\hline
\addstackgap[7pt]{(b) $f_{ni}$ \textsubscript{with $R_{p}$}}&$=\frac{1}{2{\pi}}\sqrt{\frac{g_{m}}{R_{dc}C_{dc}C_{z}}+\frac{2}{R_{p}R_{dc}C_{dc}C_{z}}}$\\\hline
\addstackgap[7pt]{(c) $f_{ni}$ \textsubscript{with $L_{p}$}}&$=\frac{1}{2{\pi}}\sqrt{\frac{g_{m}}{R_{dc}C_{dc}C_{z}}+\frac{2}{L_{p}C_{z}}}$\\\hline
\addstackgap[7pt]{(d) $f_{ni}$ \textsubscript{with $C_{p}$}}&$=\frac{1}{2{\pi}}\sqrt{\frac{g_{m}}{R_{dc}C_{dc}(C_{z}+2C_{p})}}$\\\hline
\end{tabular}\label{table:fni}
\end{table}

Although several assumptions and approximations are employed in deriving $f_{ni}$ to reduce complex calculations, the resulting equations in Table \ref{table:fni} effectively describe the frequency behavior of Type $\mathrm{II}$ in response to changes in $R_{p}$, $L_{p}$, and $C_{p}$. In Type $\mathrm{II}$ configurations shown in Figs. (9) to (12), $f_{ni}$ is proportional to the transconductance $g_{m}$, whereas $f_{ni}$ is inversely proportional to $R_{dc}$, $C_{dc}$, $C_{z}$, $R_{p}$, $L_{p}$, and $C_{p}$. Compared to Type $\mathrm{I}$ shown in Fig. \ref{fig. 5} that requires passive components to generate oscillation, Type $\mathrm{II}$ can self-oscillate by the active inductance $L_{ni}$ and active capacitance $C_{ni}$ even in the absence of passive components. Furthermore, since $R_{dc}$ and $C_{dc}$ provide independent dc bias for $M_{x}$ and $M_{y}$ in Type $\mathrm{II}$ (Fig. \ref{fig. 9}), a reliable dc bias is established. 
\section{\color{myblue}\Large{E}\large{XTENDED} \Large{A}\large{NALYSIS}}
In Section $\mathrm{III}$, to investigate the oscillation frequency $f_{ni}$ of the Type $\mathrm{II}$ configuration, the impedance analysis is conducted according to a parallel component{\textemdash}$R_{p}$, $L_{p}$, or $C_{p}${\textemdash}as summarized in Table \ref{table:fni}. \footnote{In this paper, the terms "parallel component" and "passive component" are used interchangeably to refer to $R_{p}$, $L_{p}$, and $C_{p}$.} In this section, an extended analysis of Type $\mathrm{II}$ is conducted to understand how $f_{ni}$ is affected when all parallel components are included, as depicted in Fig. \ref{fig. 1}. Recall that the series combination of $R_{ni}$, $L_{ni}$, and $C_{ni}$ depicted in Fig. \ref{fig. 1} means the equivalent circuit model of Type $\mathrm{II}$ shown in Fig. \ref{fig. 9} and Equation (26).

\subsection{\color{myblue}Type $\mathrm{II}$ with $L_{p}$ and $C_{p}$}
\begin{figure}[h!]
\centering\includegraphics[scale = 0.55]{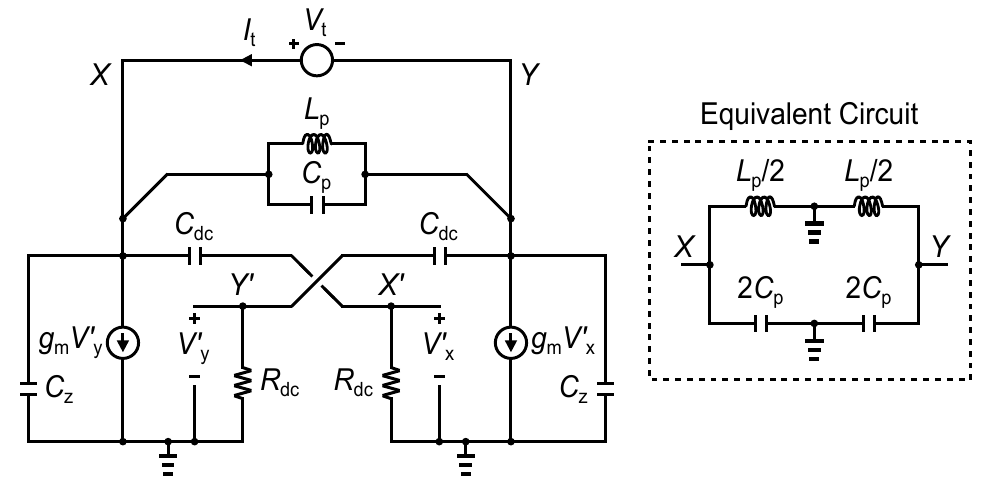}
\caption{{\color{myred}\textbf{Type $\mathrm{II}$ with $L_{p}$ and $C_{p}$.}} Small-signal model of Type $\mathrm{II}$ and an equivalent circuit of the $L_{p}$ and $C_{p}$.}\label{fig. 13}
\end{figure}

When the Type $\mathrm{II}$ configuration is combined in parallel with $L_{p}$ and $C_{p}$, its equivalent small-signal model is given by using $L_{p}/2$ and $2C_{p}$, as shown in Fig. \ref{fig. 13}, because the circuit configuration is fully differential.

In performing the small-signal analysis for Type $\mathrm{II}$ incorporating $C_{p}$ in Section $\mathrm{III}.D$, since $2C_{p}$ can merge with $C_{z}$ in Fig. \ref{fig. 12}, $f_{ni}$ for Type $\mathrm{II}$ incorporating $L_{p}$ and $C_{p}$ shown in Fig. \ref{fig. 13} can be extended from Equation (51){\textemdash}derived with only $L_{p}${\textemdash}by substituting $C_{z}+2C_{p}$ for $C_{z}$ as follows:

\begin{align}
f_{ni}=\frac{1}{2{\pi}}\sqrt{\frac{g_{m}}{R_{dc}C_{dc}(C_{z}+2C_{p})}+\frac{2}{L_{p}(C_{z}+2C_{p})}}
\end{align}

\subsection{\color{myblue}Type $\mathrm{II}$ with $R_{p}$, $L_{p}$, and $C_{p}$}
\begin{figure}[h!]
\centering\includegraphics[scale = 0.55]{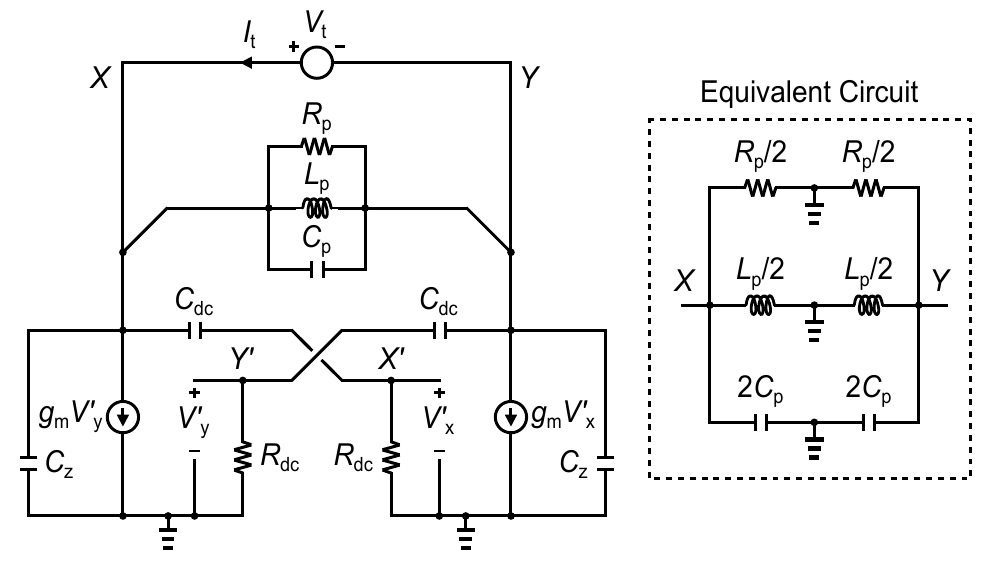}
\caption{{\color{myred}\textbf{Type $\mathrm{II}$ with $R_{p}$, $L_{p}$, and $C_{p}$.}} Small-signal model of Type $\mathrm{II}$ and an equivalent circuit of the $R_{p}$, $L_{p}$, and $C_{p}$.}\label{fig. 14}
\end{figure}

Similar to decomposing each parallel component{\textemdash}$R_{p}$, $L_{p}$, or $C_{p}${\textemdash}into two series components as performed in Figs. \ref{fig. 10}, \ref{fig. 11}, and \ref{fig. 12}, to perform the small-signal analysis for Type $\mathrm{II}$ incorporating $R_{p}$, $L_{p}$, and $C_{p}$, the parallel RLC circuit between nodes $X$ and $Y$ can be decomposed using $R_{p}/2$, $L_{p}/2$, and $2C_{p}$ as shown in Fig. \ref{fig. 14}. Recall that the equivalent circuit of Fig. \ref{fig. 14} is modeled as the circuit configuration shown in Fig. \ref{fig. 1}, and that the series combination of $R_{ni}$, $L_{ni}$, and $C_{ni}$ depicted in Fig. \ref{fig. 1} can be implemented using the cross-coupled transconductance pair incorporating $C_{z}$, $C_{dc}$, and $R_{dc}$ (Fig. \ref{fig. 14}).

In Fig. \ref{fig. 14}, applying KCL with a test voltage $V_{t}$ to nodes $X$ and $Y$ generates the following currents $I_{t}$ and $-I_{t}$:

\begin{align}
I_{t}=&g_{m}V_{y}^{\prime}+sC_{z}V_{x}+\frac{V_{x}}{Z_{dc}}\\
{\nonumber}&+\frac{V_{x}}{R_{p}/2}+\frac{V_{x}}{sL_{p}/2}+s2C_{p}V_{x}\\
=&g_{m}V_{y}^{\prime}+sC_{m}V_{x}+\frac{V_{x}}{Z_{dc}}+\frac{V_{x}}{R_{p}/2}+\frac{V_{x}}{sL_{p}/2}\\
-I_{t}=&g_{m}V_{x}^{\prime}+sC_{z}V_{y}+\frac{V_{y}}{Z_{dc}}\\
{\nonumber}&+\frac{V_{y}}{R_{p}/2}+\frac{V_{y}}{sL_{p}/2}+s2C_{p}V_{y}\\
=&g_{m}V_{x}^{\prime}+sC_{m}V_{y}+\frac{V_{y}}{Z_{dc}}+\frac{V_{y}}{R_{p}/2}+\frac{V_{y}}{sL_{p}/2}
\end{align}

{\noindent}where $C_{m}$ represents $C_{z}+2C_{p}${\textemdash}the merged capacitance at each node $X$ or $Y${\textemdash}to simplify the analysis. Recall that $V_{t}=V_{x}-V_{y}$, $V_{x}^{\prime}=V_{x}[sR_{dc}C_{dc}/(1+sR_{dc}C_{dc})]$, $V_{y}^{\prime}=V_{y}[sR_{dc}C_{dc}/(1+sR_{dc}C_{dc})]$, and $Z_{dc}=(1+sR_{dc}C_{dc})/sC_{dc}$. $V_{x}^{\prime}$ is obtained through the voltage division of $V_{x}$, it is applied to the gate terminal of $M_{y}$, and it determines the current of $M_{y}$ by $g_{m}V_{x}^{\prime}$. In the same manner as $V_{x}^{\prime}$, $V_{y}^{\prime}$ is also obtained through the voltage division of $V_{y}$, it is applied to the gate terminal of $M_{x}$, and it determines the current of $M_{x}$ by $g_{m}V_{y}^{\prime}$. $Z_{dc}$ is the series impedance of $R_{dc}$ and $C_{dc}$. Note that the small-signal model shown in Fig. \ref{fig. 14} is obtained from the Type $\mathrm{II}$ configuration shown in Fig. \ref{fig. 9}(a) except for the parallel RLC circuit.

Using the above relationships of $V_{t}$, $V_{x}^{\prime}$, $V_{y}^{\prime}$, and $Z_{dc}$, the subtraction of Equations (60) and (62) is given by

\begin{align}
2I_{t}=&-g_{m}V_{t}\frac{sR_{dc}C_{dc}}{1+sR_{dc}C_{dc}}+sC_{m}V_{t}+V_{t}\frac{sC_{dc}}{1+sR_{dc}C_{dc}}\\
{\nonumber}&+\frac{2V_{t}}{R_{p}}+\frac{2V_{t}}{sL_{p}}
\end{align}

{\noindent}Then, from Equation (63), $V_{t}/I_{t}$ is obtained as

\begin{align}
\frac{2sR_{p}L_{p}(1+sR_{dc}C_{dc})}{\left[\begin{array}{c}
-g_{m}s^2R_{p}L_{p}R_{dc}C_{dc}+s^2R_{p}L_{p}C_{m}(1+sR_{dc}C_{dc})\\
+s^2R_{p}L_{p}C_{dc}+2(1+sR_{dc}C_{dc})(R_{p}+sL_{p})
\end{array}\right]}
\end{align}

{\noindent}From Equation (64), substituting $j{\omega}$ for $s$ and rearranging the result into real and imaginary parts, $V_{t}/I_{t}$ is

\begin{align}
\frac{2R_{p}L_{p}(-{\omega}^2R_{dc}C_{dc}+j{\omega})}{\left[\begin{array}{c}
{\omega}^2g_{m}R_{p}L_{p}R_{dc}C_{dc}-{\omega}^2R_{p}L_{p}C_{m}-{\omega}^2R_{p}L_{p}C_{dc}\\
-{{\omega}^2}2L_{p}R_{dc}C_{dc}+2R_{p}\\
-j({\omega}^3R_{p}L_{p}C_{m}R_{dc}C_{dc}-{\omega}2L_{p}-{\omega}2R_{p}R_{dc}C_{dc})
\end{array}\right]}
\end{align}

{\noindent}Considering the orders of $g_{m}$, $R_{dc}$, $C_{dc}$, $R_{p}$, $L_{p}$, $C_{m}$, and ${\omega}$, the real part of the denominator in Equation (65) can be approximated as ${\omega}^2g_{m}R_{p}L_{p}R_{dc}C_{dc}$. \footnote{For example, when assuming that the Type $\mathrm{II}$ configuration shown in Fig. \ref{fig. 14} is designed based on $g_{m}=10{\,}\mathrm{mS}$, $R_{dc}=1{\,}\mathrm{k}{\Omega}$, $C_{dc}=500{\,}\mathrm{fF}$, $R_{p}=1{\,}\mathrm{M}{\Omega}$, $L_{p}=15{\,}{\mu}\mathrm{H}$, $C_{m}=200{\,}\mathrm{fF}$, and ${\omega}=2{\pi}(10^{10})$.} Therefore, $V_{t}/I_{t}$ can be simplified as

\begin{align}
\frac{2R_{p}L_{p}(-{\omega}^2R_{dc}C_{dc}+j{\omega})}{\left[\begin{array}{c}
{\omega}^2g_{m}R_{p}L_{p}R_{dc}C_{dc}\\
-j({\omega}^3R_{p}L_{p}C_{m}R_{dc}C_{dc}-{\omega}2L_{p}-{\omega}2R_{p}R_{dc}C_{dc})
\end{array}\right]}
\end{align}

To find $f_{ni}$, the reactance $X_{ni}$ is first calculated from Equation (66), and then ${\omega}_{ni}$ that makes $X_{ni}$ zero should be obtained. To obtain $X_{ni}$ from Equation (66), the complex conjugate of the denominator is multiplied by the numerator and denominator, and the numerator of $X_{ni}$ is expressed as $2R_{p}L_{p}X_{ni,num}$ as follows:

\begin{align}
2R_{p}L_{p}{\times}\underbrace{\left[\begin{array}{c}
{\omega}^3g_{m}R_{p}L_{p}R_{dc}C_{dc}\\
-{\omega}^5R_{p}L_{p}C_{m}(R_{dc}C_{dc})^2\\
+{{\omega}^3}2L_{p}R_{dc}C_{dc}+{{\omega}^3}2R_{p}(R_{dc}C_{dc})^2
\end{array}\right]}_{X_{ni,num}}
\end{align}

{\noindent}That is, $2R_{p}L_{p}X_{ni,num}$ is the numerator of the imaginary part of $V_{t}/I_{t}$. From Equation (67), solving $X_{ni,num}=0$, ${\omega}_{ni}^2$ is therefore obtained as

\begin{align}
{\omega}_{ni}^2=\frac{g_{m}R_{p}L_{p}+2L_{p}+2R_{p}R_{dc}C_{dc}}{R_{p}L_{p}C_{m}R_{dc}C_{dc}}
\end{align}

{\noindent}Applying ${\omega}_{ni}=2{\pi}f_{ni}$ and $C_{m}=C_{z}+2C_{p}$ to Equation (68), $f_{ni}${\textemdash}incorporating $R_{p}$, $L_{p}$, and $C_{p}${\textemdash}is derived as

\begin{align}
f_{ni}&=\frac{1}{2{\pi}}\sqrt{\frac{g_{m}}{R_{dc}C_{dc}C_{m}}+\frac{2}{R_{p}R_{dc}C_{dc}C_{m}}+\frac{2}{L_{p}C_{m}}}\\
&=\frac{1}{2{\pi}}\sqrt{\left(\frac{g_{m}}{R_{dc}C_{dc}}+\frac{2}{R_{p}R_{dc}C_{dc}}+\frac{2}{L_{p}}\right)\frac{1}{C_{z}+2C_{p}}}
\end{align}

{\noindent}From Equation (69), the effects of $R_{p}$ and $L_{p}$ are reflected in the second and third terms of the square root, respectively, and $C_{p}$ affects all three terms of the square root.

In Fig. \ref{fig. 14}, when choosing $R_{p}={\infty}$, $L_{p}={\infty}$, and $C_{p}=0$ to neglect the effects of parallel components, Equation (70) is modified to $f_{ni}$ in Table \ref{table:fni}(a), which is the same as the resonant frequency of the circuit configuration without parallel components as depicted in Fig. \ref{fig. 9}.

Similarly, if only $R_{p}$ is considered while neglecting $L_{p}$ and $C_{p}$ ($L_{p}={\infty}$ and $C_{p}=0$), Equation (70) is modified to $f_{ni}$ in Table \ref{table:fni}(b), representing the resonant frequency of Type $\mathrm{II}$ as depicted in Fig. \ref{fig. 10}. If only $L_{p}$ is considered while neglecting $R_{p}$ and $C_{p}$ ($R_{p}={\infty}$ and $C_{p}=0$), Equation (70) is modified to $f_{ni}$ in Table \ref{table:fni}(c), representing the resonant frequency of Type $\mathrm{II}$ as depicted in Fig. \ref{fig. 11}. If only $C_{p}$ is considered while neglecting $R_{p}$ and $L_{p}$ ($R_{p}={\infty}$ and $L_{p}={\infty}$), Equation (70) is modified to $f_{ni}$ in Table \ref{table:fni}(d), representing the resonant frequency of Type $\mathrm{II}$ as depicted in Fig. \ref{fig. 12}. As discussed above, the Type $\mathrm{II}$ configuration is capable of oscillation even without parallel (or passive) components, and can also utilize parallel (or passive) components to control its resonant frequency. 
\section{\color{myblue}\Large{L}\large{OOP} \Large{A}\large{NALYSIS}}
In Sections $\mathrm{II}$ and $\mathrm{III}$, the oscillation phenomena are analyzed using the negative impedance, which supplies oscillation energy at a resonant frequency where the reactance of the negative impedance in a series or parallel RLC circuit vanishes. Recall that Equation (3) representing the impedance of a parallel RLC circuit shows oscillation as its reactance $X_{zp}$ becomes zero, and Equation (10) representing the impedance of a series RLC circuit shows oscillation as its reactance $X_{ni}$ becomes zero.

In this section, the oscillation phenomena of Types $\mathrm{I}$ and $\mathrm{II}$ are revisited using the Barkhausen criteria{\textemdash}a loop gain is equal to or greater than unity, and the total phase shift of a loop is an integer multiple of $2{\pi}$ to generate oscillation.

\subsection{\color{myblue}Frequency Response of Type $\mathrm{I}$}
\begin{figure}[h!]
\centering\includegraphics[scale = 0.55]{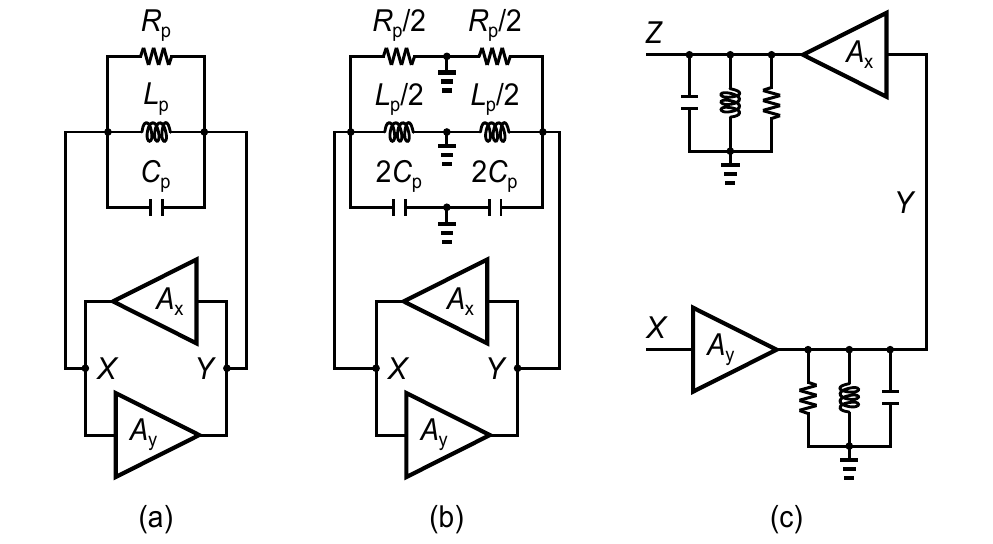}
\caption{{\color{myred}\textbf{Type $\mathrm{I}$ with a parallel RLC circuit.}} (a) Equivalent model. (b) Equivalent decomposition of the parallel RLC circuit. (c) Loop decomposition.}\label{fig. 15}
\end{figure}

Considering the Type $\mathrm{I}$ configuration with a parallel RLC circuit, the configuration can be redrawn as Fig. \ref{fig. 15}(a). The amplifiers $A_{x}$ and $A_{y}$ represent the transistors $M_{x}$ and $M_{y}$ in Fig. \ref{fig. 5}(a), respectively, to implement a cross-coupled transconductance pair. Since $A_{x}$ and $A_{y}$ are designed differentially at the transistor level, each parallel component can be decomposed into two series components separated by a virtual ground (Fig. \ref{fig. 15}(b)). Then, breaking the loop formed by $A_{x}$ and $A_{y}$ (Fig. \ref{fig. 15}(c)), the loop configuration can be decomposed into two amplification stages{\textemdash}the first stage is defined from node $X$ to node $Y$, and the second stage is defined from node $Y$ to node $Z$. Consequently, the outputs of $A_{x}$ and $A_{y}$ are modeled using the equivalent circuit consisting of $R_{p}/2$, $L_{p}/2$, and $2C_{p}$.

Recall that the voltage gain of a single transistor, $M_{x}$ or $M_{y}$, is defined as $-g_{m}r_{o}${\textemdash}the product of $g_{m}$ and the "output resistance". \footnote{The voltage gain of a single transistor is derived in Appendix $\mathrm{A.1}$.} In Fig. \ref{fig. 15}(c), the voltage gain from node $X$ to node $Z$ is defined using two consecutive stages{\textemdash}$V_{y}/V_{x}$ and $V_{z}/V_{y}${\textemdash}as follows:

\begin{align}
\frac{V_{z}}{V_{x}}=\frac{V_{y}}{V_{x}}\frac{V_{z}}{V_{y}}
\end{align}

In Fig. \ref{fig. 15}(c), assuming that the transconductance of $A_{x}$ is the same as that of $A_{y}${\textemdash}denoted as $g_{m}${\textemdash}and defining the resistance (also called the "real part") of an equivalent parallel RLC circuit is $R_{zp}^{\prime\prime}$, the gains $V_{y}/V_{x}$ and $V_{z}/V_{y}$ are expressed identically as $-g_{m}(r_{o}||R_{zp}^{\prime\prime})$. That is,

\begin{align}
-g_{m}\left[r_{o}||\underbrace{\frac{{\omega}^2\frac{R_{p}}{2}\left(\frac{L_{p}}{2}\right)^2}{\left(\frac{R_{p}}{2}\right)^2\left(1-{\omega}^2L_{p}C_{p}\right)^2+{\omega}^2\left(\frac{L_{p}}{2}\right)^2}}_{R_{zp}^{\prime\prime}}\right]
\end{align}

{\noindent}where $R_{zp}^{\prime\prime}$ is obtained by scaling $R_{p}$, $L_{p}$, and $C_{p}$ contained in $R_{zp}$ of Equation (3). When decomposing the parallel RLC circuit into the two equivalent parallel RLC circuits (Figs. \ref{fig. 15}(a) and (b)), $R_{p}$ and $L_{p}$ are scaled down by half, and $C_{p}$ increases by a factor of two. Accordingly, $R_{zp}^{\prime\prime}$ is derived from $R_{zp}$ of Equation (3).

From Equation (72), by setting the output resistance $r_{o}$ to be larger than $R_{zp}^{\prime\prime}$, the gain $-g_{m}(r_{o}||R_{zp}^{\prime\prime})$ is approximated as $-g_{m}R_{zp}^{\prime\prime}$. The voltage gain from node $X$ to node $Z$ is therefore given by

\begin{align}
\frac{V_{z}}{V_{x}}=g_{m}^2\left[\frac{{\omega}^2\frac{R_{p}}{2}\left(\frac{L_{p}}{2}\right)^2}{\left(\frac{R_{p}}{2}\right)^2\left(1-{\omega}^2L_{p}C_{p}\right)^2+{\omega}^2\left(\frac{L_{p}}{2}\right)^2}\right]^2
\end{align}

Assuming that a signal travels around the loop depicted in Fig. \ref{fig. 15}(c) and considering the gain in Equation (73), the signal polarity of node $X$ is equal to that of node $Z$, and the gain is maximized at a frequency $1/(2{\pi}\sqrt{L_{p}C_{p}})$.

Consequently, connecting nodes $X$ and $Z$ establishes a positive feedback loop driven by the same polarities at these two nodes, and the signal at $1/(2{\pi}\sqrt{L_{p}C_{p}})${\textemdash}where the maximum gain occurs{\textemdash}dominates this feedback loop, triggering oscillation at the frequency of $1/(2{\pi}\sqrt{L_{p}C_{p}})$.

\subsection{\color{myblue}Frequency Response of Type $\mathrm{II}$}
\begin{figure}[h!]
\centering\includegraphics[scale = 0.55]{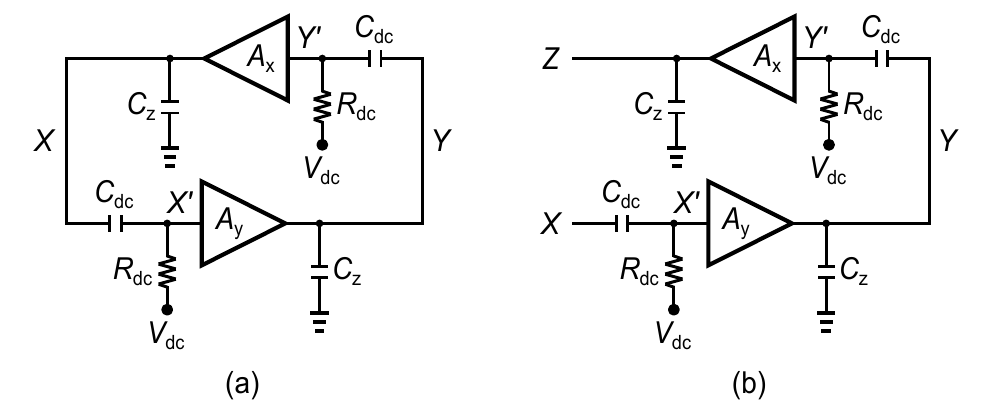}
\caption{{\color{myred}\textbf{Oscillator based on Type $\mathrm{II}$.}} (a) Equivalent model. (b) Loop decomposition.}\label{fig. 16}
\end{figure}

Fig. \ref{fig. 16}(a) describes the equivalent model of Type $\mathrm{II}$ shown in Fig. \ref{fig. 9}(a). The amplifiers $A_{x}$ and $A_{y}$ represent the transistors $M_{x}$ and $M_{y}$ shown in Fig. \ref{fig. 9}(a), respectively, and establish a cross-coupled transconductance pair. $R_{dc}$, $C_{dc}$, and $V_{dc}$ establish independent dc biases for $A_{x}$ and $A_{y}$ through nodes $Y^{\prime}$ and $X^{\prime}$, respectively. $V_{dc}$ is delivered to the input terminals of $A_{x}$ and $A_{y}$ through $R_{dc}$, setting their dc biases and determining their bandwidths. $C_{z}$ also controls the bandwidths of $A_{x}$ and $A_{y}$. The bandwidth variation by $V_{dc}$ and $C_{z}$ contributes to the change in the oscillation frequency of Type $\mathrm{II}$.

To analyze the gain and phase shift of Type $\mathrm{II}$ and ultimately understand the oscillation characteristics, the loop of the cross-coupled transconductance pair is broken down into the open loop shown in Fig. \ref{fig. 16}(b). By breaking node $X$ in Fig. \ref{fig. 16}(a), the output of $A_{x}$ is defined as node $Z$ as shown in Fig. \ref{fig. 16}(b). Assuming that an arbitrary input signal is applied to node $X$ and travels to node $Z$ in Fig. \ref{fig. 16}(b), the signal experiences the four stages through nodes $X$, $X^{\prime}$, $Y$, $Y^{\prime}$, and $Z$ as follows:

\begin{align}
\frac{V_{z}}{V_{x}}=\frac{V_{x^{\prime}}}{V_{x}}\frac{V_{y}}{V_{x^{\prime}}}\frac{V_{y^{\prime}}}{V_{y}}\frac{V_{z}}{V_{y^{\prime}}}
\end{align}

{\noindent}where $V_{x^{\prime}}/V_{x}$ and $V_{y^{\prime}}/V_{y}$ represent the transfer functions of the high-pass filters formed by $R_{dc}$ and $C_{dc}$. The transfer function of each high-pass filter is expressed as follows:

\begin{align}
\frac{sR_{dc}C_{dc}}{1+sR_{dc}C_{dc}}&=\frac{s}{\frac{1}{R_{dc}C_{dc}}+s}\\
&=\frac{s}{{\omega}_{p,RC}+s}
\end{align}

{\noindent}where a pole is at ${\omega}_{p,RC}=1/(R_{dc}C_{dc})$, and a zero is at ${\omega}_{z,RC}=0$. Therefore, the pole and zero frequencies are $f_{p,RC}=1/(2{\pi}R_{dc}C_{dc})$ and $f_{z,RC}=0$, respectively.

In Fig. \ref{fig. 16}(b), assuming that the transconductances of $A_{x}$ and $A_{y}$ are equal to $g_{m}$, and the output resistances of $A_{x}$ and $A_{y}$ are equal to $r_{o}$, the transfer functions of $V_{y}/V_{x^{\prime}}$ and $V_{z}/V_{y^{\prime}}$ are the same as

\begin{align}
\frac{-g_{m}r_{o}}{1+\frac{s}{{\omega}_{p,A}}}=\frac{-g_{m}r_{o}{\omega}_{p,A}}{{\omega}_{p,A}+s}
\end{align}

{\noindent}where ${\omega}_{p,A}$ is a pole expressed as $1/(r_{o}C_{z})$, and the pole frequency is $f_{p,A}=1/(2{\pi}r_{o}C_{z})$. From Equations (76) and (77), the transfer functions of $V_{y}/V_{x}$ and $V_{z}/V_{y}$ are equal to

\begin{align}
\frac{-g_{m}r_{o}{\omega}_{p,A}s}{({\omega}_{p,RC}+s)({\omega}_{p,A}+s)}
\end{align} 

{\noindent}where two pole frequencies are obtained as $f_{p,RC}$ and $f_{p,A}$, one zero frequency is obtained as $f_{z,RC}$, and the phase difference between the input and output is $-180^{\circ}$. Note that since the design parameters from node $X$ to node $Y$ are identical to those from node $Y$ to node $Z$, $V_{x^{\prime}}/V_{x}=V_{y^{\prime}}/V_{y}$, $V_{y}/V_{x^{\prime}}=V_{z}/V_{y^{\prime}}$, and $V_{y}/V_{x}=V_{z}/V_{y}$.

To analyze the gain and phase shift from node $X$ to node $Y$ for two cases $f_{p,A}<f_{p,RC}$ and $f_{p,A}>f_{p,RC}$, the Bode plots of the transfer functions{\textemdash}$V_{x^{\prime}}/V_{x}$, $V_{y}/V_{x^{\prime}}$, and $V_{y}/V_{x}${\textemdash}are summarized in Fig. \ref{fig. 17}. \footnote{The basic principles of the Bode plot are described in Appendix $\mathrm{A.2}$.} The magnitude and phase shift of $V_{x^{\prime}}/V_{x}$ are denoted by $|X^{\prime}/X|$ and ${\angle}X^{\prime}/X$, respectively (Fig. \ref{fig. 17}). In the same way, the magnitudes of $V_{y}/V_{x^{\prime}}$ and $V_{y}/V_{x}$ are respectively denoted by $|Y/X^{\prime}|$ and $|Y/X|$, and the phase shifts are respectively denoted by ${\angle}Y/X^{\prime}$ and ${\angle}Y/X$.

\begin{figure}[h!]
\centering\includegraphics[scale = 0.55]{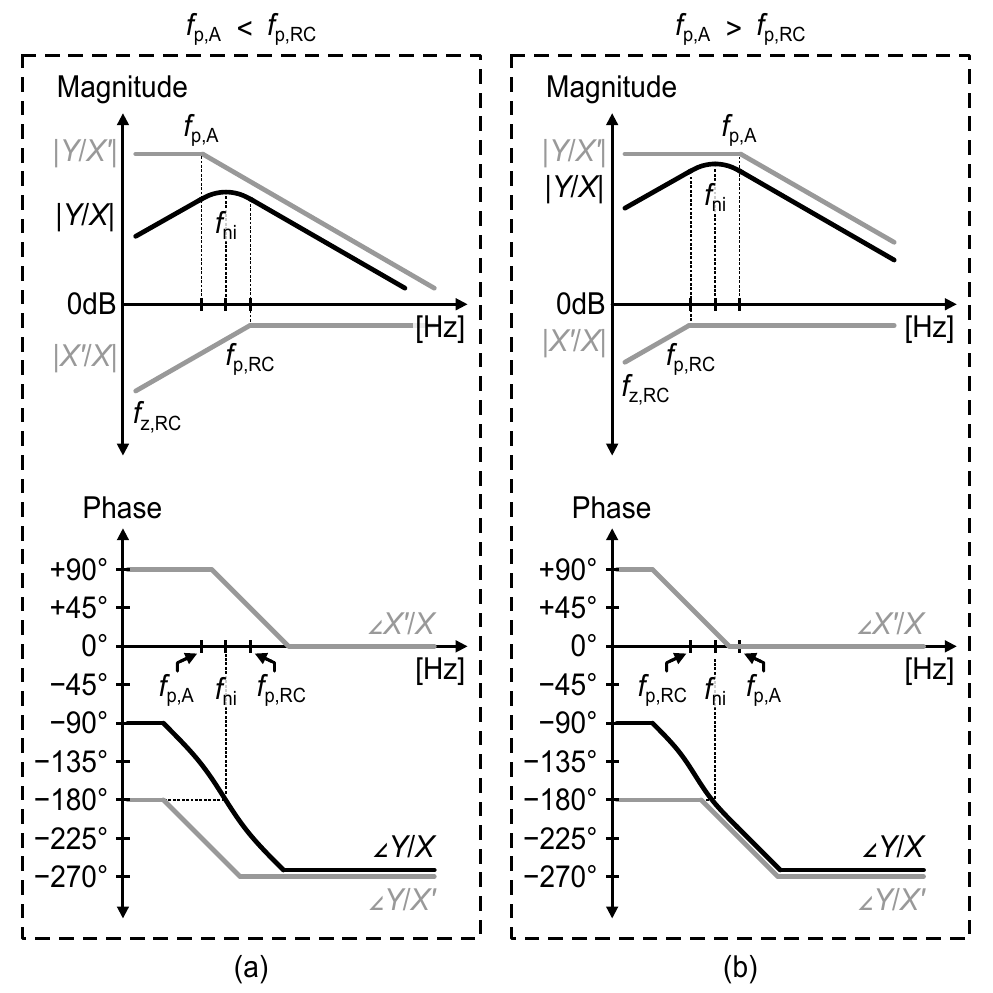}
\caption{{\color{myred}\textbf{Bode plots of Type $\mathrm{II}$ from node $X$ to node $Y$.}} (a) When $f_{p,A}<f_{p,RC}$. (b) When $f_{p,A}>f_{p,RC}$.}\label{fig. 17}
\end{figure}

From Equation (76), since $V_{x^{\prime}}/V_{x}$ consists of a zero at $f_{z,RC}=0$ and a pole at $f_{p,RC}=1/(2{\pi}R_{dc}C_{dc})$, the magnitude $|X^{\prime}/X|$ increases with a slope of $+20{\,}\mathrm{dB/decade}$ starting from $f_{z,RC}$, and then flattens out after passing $f_{p,RC}$ which contributes to $-20{\,}\mathrm{dB/decade}$. After passing $f_{p,RC}$, $|X^{\prime}/X|$ approaches $0{\,}\mathrm{dB}$, forming a high-pass filter. In the phase domain, ${\angle}X^{\prime}/X$ shows the phase contribution starting from $0.1f_{z,RC}$, reaches a contribution of $+45^{\circ}$ at $f_{z,RC}$, and approaches a contribution of $+90^{\circ}$ at $10f_{z,RC}$. However, $f_{z,RC}$ is formed at $0{\,}\mathrm{Hz}$, implying that the observation of a signal must be performed for an infinite time, and this cannot be achieved in reality. In other words, even though $f_{z,RC}$ is theoretically $0$, it must be treated as a very low frequency, such as less than $1\mathrm{Hz}$, for the analysis. Therefore, due to the unreachable and unmeasurable $f_{z,RC}$, ${\angle}X^{\prime}/X$ has an initial phase shift of approximately $+90^{\circ}$ immediately after $10f_{z,RC}$ (Fig. \ref{fig. 17}). Then, ${\angle}X^{\prime}/X$ begins the phase contribution from the initial phase shift of $+90^{\circ}$ at $0.1f_{p,RC}$, reaches $+45^{\circ}$ by a $-45^{\circ}$ contribution at $f_{p,RC}$, and approaches $0^{\circ}$ by a $-90^{\circ}$ contribution at $10f_{p,RC}$.

From Equation (77), $V_{y}/V_{x^{\prime}}$ exhibits a gain of $-g_{m}r_{o}$ at low frequencies and has a pole at $f_{p,A}=1/(2{\pi}r_{o}C_{z})$. Therefore, the magnitude $|Y/X^{\prime}|$ remains flat (or $g_{m}r_{o}$) to $f_{p,A}$ and decreases with a slope of $-20{\,}\mathrm{dB/decade}$ after passing $f_{p,A}$. Since the sign of $V_{y}/V_{x^{\prime}}$ is negative, the initial phase shift of ${\angle}Y/X^{\prime}$ begins at $-180^{\circ}$ (Fig. \ref{fig. 17}). Subsequently, as the frequency increases from a very low value, ${\angle}Y/X^{\prime}$ remains $-180^{\circ}$ in the low-frequency band, begins the phase contribution at $0.1f_{p,A}$, reaches $-225^{\circ}$ by a $-45^{\circ}$ contribution at $f_{p,A}$, and approaches $-270^{\circ}$ by a $-90^{\circ}$ contribution at $10f_{p,A}$. The phase plots in Fig. \ref{fig. 17} show that the initial phase shifts of ${\angle}X^{\prime}/X$ and ${\angle}Y/X^{\prime}$ are set to $+90^{\circ}$ by $f_{z,RC}=0$ and to $-180^{\circ}$ by $-g_{m}r_{o}$, respectively.

Based on the above observations, the two consecutive responses{\textemdash}$|X^{\prime}/X|$ and $|Y/X^{\prime}|$; ${\angle}X^{\prime}/X$ and ${\angle}Y/X^{\prime}${\textemdash}are combined to obtain $|Y/X|$ and ${\angle}Y/X$. From Equation (78), as the frequency changes from a very low value to a high value in Fig. \ref{fig. 17}, $|Y/X|$ initially increases with a slope of $+20{\,}\mathrm{dB/decade}$ in the low-frequency band due to $f_{z,RC}=0$, and subsequently reaches a maximum magnitude at $f_{ni}$ by the two consecutive poles, $f_{p,RC}$ and $f_{p,A}$. Then, after passing a second pole, $|Y/X|$ decreases with a slope of $-20{\,}\mathrm{dB/decade}$. As a result, $|Y/X|$ forms a resonant filter with a peak frequency of $f_{ni}$.

At very low frequencies, the initial phase shift of ${\angle}Y/X$ is formed as approximately $-90^{\circ}$ by combining $+90^{\circ}$ for ${\angle}X^{\prime}/X$ and $-180^{\circ}$ for ${\angle}Y/X^{\prime}$ (Fig. \ref{fig. 17}). Then, as the frequency increases, ${\angle}Y/X$ reaches $-135^{\circ}$ and $-225^{\circ}$ at the two poles, respectively, and ultimately approaches $-270^{\circ}$ after passing $f_{p,RC}$ and $f_{p,A}$. As ${\angle}Y/X$ varies from $-90^{\circ}$ to $-270^{\circ}$, a phase shift of $-180^{\circ}$ is achieved at $f_{ni}$ formed between the two poles. In other words, $V_{y}/V_{x}$ exhibits its maximum gain and a phase shift of $-180^{\circ}$ at $f_{ni}$ formed between $f_{p,RC}$ and $f_{p,A}$.

In the decomposed loop depicted in Fig. \ref{fig. 16}(b), when a signal travels from node $X$ to node $Z$, the maximum gain is achieved at $f_{ni}$, and the phase shift becomes $-360^{\circ}$ (or $2{\pi}$) at $f_{ni}$. Consequently, when the loop is restored by connecting nodes $X$ and $Z$, a positive feedback loop is established, leading to oscillation at the frequency $f_{ni}$.

\begin{figure}[h!]
\centering\includegraphics[scale = 0.55]{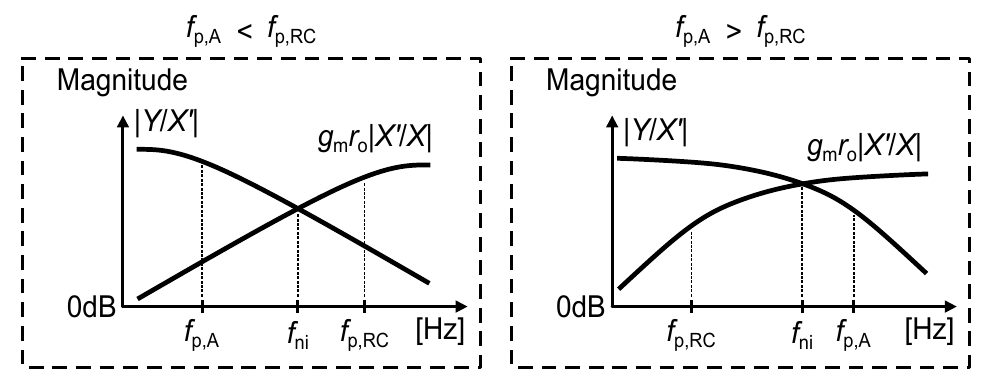}
\caption{{\color{myred}\textbf{Actual magnitudes of $g_{m}r_{o}|X^{\prime}/X|$ and $|Y/X^{\prime}|$.}} Intersection points for the two cases.}\label{fig. 18}
\end{figure}

As an intuitive way, $f_{ni}$ can be approximated by finding an intersection point of the two magnitudes, $g_{m}r_{o}|X^{\prime}/X|$ and $|Y/X^{\prime}|$. In the magnitude plots shown in Figs. \ref{fig. 17}(a) and (b), the oscillation frequency $f_{ni}$ is formed at a point where $|Y/X^{\prime}|$ intersects with the high-pass filter formed by $g_{m}r_{o}|X^{\prime}/X|$ having the same gain as the low-frequency gain of $|Y/X^{\prime}|$. Note that the magnitudes of $|X^{\prime}/X|$ and $|Y/X^{\prime}|$ are denoted by straight lines in Fig. \ref{fig. 17} to simplify the analysis, but the actual magnitudes change gradually in the vicinity of poles. This means that the magnitudes are represented by curved lines as depicted in Fig. \ref{fig. 18}. Also, the intersection point formed by $g_{m}r_{o}|X^{\prime}/X|$ and $|Y/X^{\prime}|$ is represented as $f_{ni}$ for both cases, $f_{p,A}<f_{p,RC}$ and $f_{p,A}>f_{p,RC}$, in Fig. \ref{fig. 18}. That is, the respective movements of $f_{p,RC}$ and $f_{p,A}$ provide an intuitive grasp of the shift in $f_{ni}$. By modifying Equations (76) and (77), the intersection point can be found using the following equation

\begin{align}
\underbrace{\frac{g_{m}r_{o}s}{{\omega}_{p,RC}+s}}_{g_{m}r_{o}|X^{\prime}/X|}=\underbrace{\frac{g_{m}r_{o}{\omega}_{p,A}}{{\omega}_{p,A}+s}}_{|Y/X^{\prime}|}
\end{align}

{\noindent}Equation (79) is rearranged as follows:

\begin{align}
g_{m}r_{o}s({\omega}_{p,A}+s)&=g_{m}r_{o}{\omega}_{p,A}({\omega}_{p,RC}+s)\\
s^2+{\omega}_{p,A}s&={\omega}_{p,A}s+{\omega}_{p,RC}{\omega}_{p,A}\\
s^2&={\omega}_{p,RC}{\omega}_{p,A}
\end{align}

{\noindent}where ${\omega}_{p,RC}=1/(R_{dc}C_{dc})$ and ${\omega}_{p,A}=1/(r_{o}C_{z})$. From Equations (79) and (82), the intersection point{\textemdash}the two magnitudes of $g_{m}r_{o}|X^{\prime}/X|$ and $|Y/X^{\prime}|$ become equal as depicted in Fig. \ref{fig. 18}{\textemdash}is determined by ${\omega}=\sqrt{{\omega}_{p,RC}{\omega}_{p,A}}$. Therefore, ${\omega}_{ni}$ and $f_{ni}$ are obtained as follows:

\begin{align}
{\omega}_{ni}&=\sqrt{\frac{1}{R_{dc}C_{dc}}\frac{1}{r_{o}C_{z}}}=\sqrt{\frac{g_{m}}{R_{dc}C_{dc}C_{z}}\frac{{\lambda}V_{ov}}{2}}\\
f_{ni}&=\frac{1}{2{\pi}}\sqrt{\frac{g_{m}}{R_{dc}C_{dc}C_{z}}\frac{{\lambda}V_{ov}}{2}}
\end{align}

{\noindent}where the output resistance $r_{o}$ is $2/({\lambda}g_{m}V_{ov})$. \footnote{The transconductance $g_{m}$ of a CMOS transistor is defined as $2I_{D}/V_{ov}$, the overdrive voltage $V_{ov}$ is obtained as $V_{gs}-V_{th}$, the drain current $I_{D}$ can be rearranged as $g_{m}V_{ov}/2$, and the output resistance $r_{o}$ of a CMOS transistor is defined as $1/({\lambda}I_{D})$. Using $I_{D}=g_{m}V_{ov}/2$ and the channel-length modulation coefficient ${\lambda}$, $r_{o}$ can be modified to $2/({\lambda}g_{m}V_{ov})$. The details of the CMOS parameters are described in \cite{Razavi2017parameter}.} Similar to the equation of $f_{ni}=[1/(2{\pi})]\sqrt{g_{m}/(R_{dc}C_{dc}C_{z})}$ derived from Fig. \ref{fig. 9}(b), Equation (84) exhibits frequency behaviors that are proportional to $g_{m}$ and inversely proportional to $R_{dc}$, $C_{dc}$, and $C_{z}$. However, while the output resistance $r_{o}$ is assumed as an infinite value during the derivation of $f_{ni}$ in Fig. \ref{fig. 9}(b), $r_{o}$ is used as a finite value when deriving $f_{ni}$ from the open loop shown in Fig. \ref{fig. 16}(b). As a result, the channel-length modulation coefficient ${\lambda}$ and overdrive voltage $V_{ov}$ are included in the expression of $f_{ni}$. 
\section{\color{myblue}\Large{C}\large{ONCLUSION}}
In this work, oscillation phenomena are analyzed employing the configurations of Types $\mathrm{I}$ and $\mathrm{II}$. In Section $\mathrm{I}$, the basic principles of parallel and series RLC circuits are discussed. In Sections $\mathrm{II}$, $\mathrm{III}$, and $\mathrm{IV}$, the oscillation mechanisms based on the negative impedances{\textemdash}for Types $\mathrm{I}$ and $\mathrm{II}${\textemdash}are discussed through the small-signal analysis. In Section $\mathrm{V}$, the transfer function analysis for Types $\mathrm{I}$ and $\mathrm{II}$ is conducted to understand oscillation mechanisms.

The negative impedance analysis for Types $\mathrm{I}$ and $\mathrm{II}$, conducted in Sections $\mathrm{II}$, $\mathrm{III}$, and $\mathrm{IV}$, demonstrates how the design parameters{\textemdash}$g_{m}$, $R_{dc}$, $C_{dc}$, and $C_{z}${\textemdash}affect the oscillation frequency $f_{ni}$ in the presence and absence of passive components{\textemdash}$R_{p}$, $L_{p}$, and $C_{p}$.

The transfer function analysis for Types $\mathrm{I}$ and $\mathrm{II}$, conducted in Section $\mathrm{V}$, enables an intuitive understanding of the changes in $f_{ni}$ by converting the closed-loop system into the open-loop system.

In summary, Types $\mathrm{I}$ and $\mathrm{II}$ oscillate at the frequency $f_{ni}$ determined by passive components{\textemdash}$R_{p}$, $L_{p}$, and $C_{p}$. In particular, Type $\mathrm{II}$ not only operates in conjunction with the passive components, but also can oscillate without any passive components by utilizing its internally generated negative inductance and capacitance{\textemdash}$L_{ni}$ and $C_{ni}$. 
\pretocmd{\thebibliography}{\color{mygreen}}{}{}
\balance

\balance

\newpage
\section{\color{myblue}\Large{A}\large{PPENDIX}}
\renewcommand{\thefigure}{S\arabic{figure}}
\setcounter{figure}{0}

\begin{figure}[h!]
\centering\includegraphics[scale = 0.55]{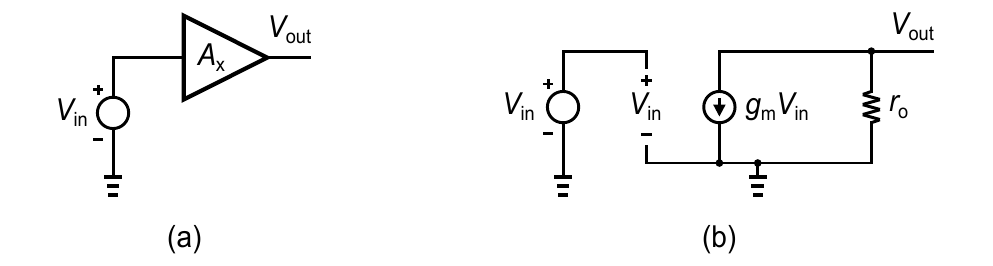}
\caption{{\color{myred}\textbf{Amplification model.}} (a) Amplifier $A_{x}$. (b) Its small-signal model.}\label{fig. S1}
\end{figure}

\noindent\textbf{\color{myblue}\textit{A}.1.\space{\,}Voltage Gain of a Single Transistor}\\
{\indent}In Fig. \ref{fig. S1}(a), the amplifier $A_{x}$ represents a common-source stage of a CMOS transistor with an input $V_{in}$ and an output $V_{out}$. Fig. \ref{fig. S1}(b) describes the small-signal model of $A_{x}$ with an infinite input impedance and a finite output impedance $r_{o}$. $g_{m}$ means the transconductance of $A_{x}$, and the output stage of $A_{x}$ draws the resulting current $g_{m}V_{in}$ in the small-signal model. To simplify the small-signal analysis, parasitic components between the input, output, and ground are neglected.

Applying $V_{in}$ to the input terminal of $A_{x}$, the KCL result at the output terminal is obtained as follows:

\begin{align}
g_{m}V_{in}+\frac{V_{out}-0}{r_{o}}=0
\end{align}

The voltage gain $V_{out}/V_{in}$ is therefore derived as $-g_{m}r_{o}$. Since $V_{out}/V_{in}$ has the negative sign, $V_{in}$ is amplified with a gain of $g_{m}r_{o}$, and $V_{out}$ exhibits a $180^{\circ}$ phase shift relative to $V_{in}$. In other words, the voltage gain is defined as the product of $g_{m}$ and the "output resistance" seen between the output terminal and ground.{\singlespacing}

\begin{figure}[h!]
\centering\includegraphics[scale = 0.55]{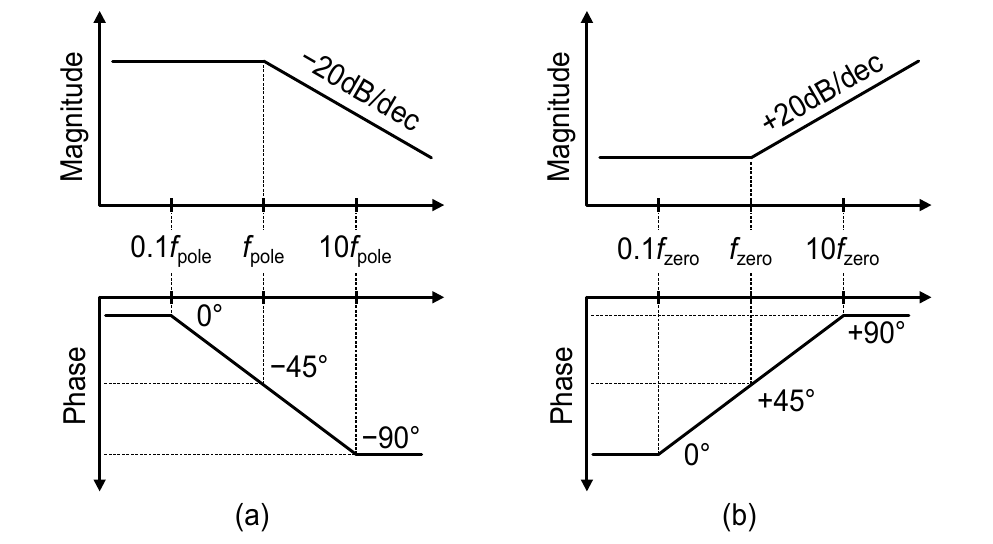}
\caption{{\color{myred}\textbf{Bode plot.}} (a) Pole. (b) Zero.}\label{fig. S2}
\end{figure}

\noindent\textbf{\color{myblue}\textit{A}.2.\space{\,}Properties of Poles and Zeros}\\
{\indent}In the Bode plot depicted in Fig. \ref{fig. S2}(a), when a system has a pole $f_{pole}$, the magnitude of the gain begins to decrease with a slope of $-20{\,}\mathrm{dB/decade}$ starting from $f_{pole}$. From the perspective of phase shift, the system exhibits a phase shift across the frequency band of $0.1f_{pole}$ through $10f_{pole}$. The system begins contributing to the phase shift at $0.1f_{pole}$, reaches a contribution of $-45^{\circ}$ at $f_{pole}$, and approaches a contribution of $-90^{\circ}$ at $10f_{pole}$.

In the Bode plot depicted in Fig. \ref{fig. S2}(b), when a system encounters a zero $f_{zero}$, it exhibits different behaviors compared to when the system encounters $f_{pole}$. When the system encounters $f_{zero}$, the magnitude of its gain begins to increase with a slope of $+20{\,}\mathrm{dB/decade}$. From the perspective of phase shift, the system exhibits a phase shift across the frequency band of $0.1f_{zero}$ through $10f_{zero}$. The system begins contributing to the phase shift at $0.1f_{zero}$, reaches a contribution of $+45^{\circ}$ at $f_{zero}$, and approaches a contribution of $+90^{\circ}$ at $10f_{zero}$.

In the phase shifts depicted in Figs. \ref{fig. S2}(a) and (b), the system does not "reach" $-90^{\circ}$ at $10f_{pole}$, but "approaches" $-90^{\circ}$, strictly speaking. Similarly, at $10f_{zero}$, the system "approaches" $+90^{\circ}$ rather than "reaching" $+90^{\circ}$. Only when a frequency far exceeds beyond $10f_{pole}$ (or $10f_{zero}$) and becomes infinite, the system reaches a phase contribution of $-90^{\circ}$ (or $+90^{\circ}$) after a pole (or zero).

If $f_{pole}$ and $f_{zero}$ are closely located within a system: (1) the system bandwidth is extended because $+20{\,}\mathrm{dB/decade}$ by $f_{zero}$ suppresses $-20{\,}\mathrm{dB/decade}$ by $f_{pole}$; (2) the phase shifts of $f_{pole}$ and $f_{zero}$ suppress each other, resulting in a net phase contribution of $0^{\circ}$ and stabilizing a negative feedback system. That is, $f_{zero}$ compensates for $f_{pole}$, extending the bandwidth and improving the phase margin.

According to circuit architectures requiring negative or positive feedback, such as amplifiers or oscillators, poles and zeros must be appropriately managed by adding or suppressing them as needed to ensure the performance and reliability. 
\end{document}